\documentclass[12pt]{amsart}
\newcommand{\sumprime}{\mathop{{\sum}'}}
\usepackage{amsfonts,amssymb}
\title[Laplace transformations and spectral theory]{Laplace transformations and
spectral theory of
two-dimensional semi-discrete and discrete hyperbolic Schr\"odinger operators}
\author{Alexei~A.~Oblomkov}
\address{Department of Mathematics, Massachusetts Institute of Technology,
77 Massachusetts Ave., Cambridge, MA 02139, USA.}
\email{oblomkov@math.mit.edu}
\author{Alexei~V.~Penskoi}
\address{Centre de Recherches Math\'ematiques, Universit\'e de Montr\'eal,
C.~P.~6128, Succ. Centre-ville, Montr\'eal, 
Qu\'ebec, H3C 3J7, Canada.} 
\email{penskoi@crm.umontreal.ca}
\date{}
\newtheorem{Theorem}{Theorem}
\newtheorem{Lemma}{Lemma}
\newtheorem{Definition}{Definition}
\begin{document}
\begin{abstract}
We introduce Laplace transformations of 2D semi-discrete
hyperbolic Schr\"odinger operators and show their relation to a semi-discrete 
2D Toda lattice. We develop the algebro-geometric spectral theory of 2D 
semi-discrete hyperbolic Schr\"odinger operators and solve the direct 
spectral problem for 2D discrete ones 
(the inverse problem for discrete operators was already solved by Krichever). 
Using the spectral theory we investigate spectral 
properties of the Laplace transformations of these operators.
This makes it possible 
to find solutions of the semi-discrete and discrete 2D
Toda lattices in terms of theta-functions.
\end{abstract}

\maketitle

\section{Introduction}

The interest in the transformations
$$
L=\frac{1}{2}(\partial_x+A)(\partial_y+B)+W \mapsto%
\tilde{L}=\frac{1}{2}W(\partial_y+B)W^{-1}(\partial_x+A)+W,  
$$
$$
L=\frac{1}{2}(\partial_y+B)(\partial_x+A)+V \mapsto%
\hat{L}=\frac{1}{2}V(\partial_x+A)V^{-1}(\partial_y+B)+V  
$$
of the two-dimensional hyperbolic Schr\"odinger operator 
$L=\frac{1}{2}\partial_x\partial_y+F\partial_x+G\partial_y+H$
goes back to
Laplace. These transformations act also on the solutions of 
the equation $L\psi=0:$
$$
L\mapsto\tilde{L},\quad \psi\mapsto\tilde{\psi}=(\partial_y+B)\psi,
$$
$$
L\mapsto\hat{L},\quad \psi\mapsto\hat{\psi}=(\partial_x+A)\psi.
$$
The Laplace transformations are useful in the theory of congruences
of surfaces in $\mathbb{R}^3$ and they were studied by Darboux, Tzitz\'eica 
and others (references and a more extended exposition can be found in 
the paper~\cite{ND}). It was remarked already then that the 
chain of Laplace transformations $\dots,L_{-1},L_0,L_1,\dots,$
where $L_{i+1}=\tilde{L}_i,$ is equivalent to the non-linear equation 
\begin{equation}\label{2DToda}
\frac{1}{2}\partial_x\partial_yg_n=e^{g_{n+1}-g_n}-e^{g_n-g_{n-1}}
\end{equation}
now called the 2D Toda lattice. Its integrability from different points of view
was discovered by Mikhailov~\cite{M}, Fordy and Gibbons~\cite{FG},
Leznov and Savel'ev~\cite{LS}, 
Bulgadaev~\cite{B}.

Various generalizations of the Laplace transformations were also studied
(a review can be found in the paper~\cite{ND}), among them
the Laplace transformations of two-dimensional elliptic Schr\"odinger 
operators 
$$
L=\frac{1}{2}(\bar{\partial}+B)(\partial+A)+V,\quad 
\partial=\partial_x-i\partial_y,
$$
and the Laplace transformations of discrete hyperbolic Schr\"odinger 
operators
\begin{equation}\label{discrete}
(L\psi)_{n,m}=a_{n,m}\psi_{n,m}+b_{n,m}\psi_{n+1,m}+c_{n,m}\psi_{n,m+1}+%
d_{n,m}\psi_{n+1,m+1}.
\end{equation}
In both cases chains of Laplace transformations are related to the
corresponding versions of the 2D Toda lattice. In the case of elliptic 
Schr\"odinger operators one of the principal results concerns the description of
cyclic chains of Laplace transformations 
(such that $L_N=L_0$ for some $N$). 
It was proven by Novikov and Veselov~\cite{NV1,NV2} that if we consider
cyclic chains of Laplace transformations of {\em periodic} elliptic 
Schr\"odinger operators 
then the operators in such chains are topologically trivial algebro-geometric 
operators.  

These results were the motivation for this paper. Our goal is to study 
Laplace transformations of two-dimensional semi-discrete 
\begin{equation}\label{semi-discrete}
(L\psi)_n=a_n(y)\psi_n(y)+b_n(y)\psi'_n(y)+c_n(y)\psi_{n+1}(y)+%
d_n(y)\psi'_{n+1}(y)
\end{equation}
and discrete~(\ref{discrete}) hyperbolic Schr\"odinger operators. 
To the best of the authors' knowledge, the Laplace transformations of 
semi-discrete operators were not yet studied. We start by introducing 
Laplace transformations of semi-discrete
Schr\"odinger operators and showing their relation to a semi-discrete 2D Toda 
lattice. Then we study spectral properties of the Laplace 
transformation of {\em periodic} operators.

Let us recall that the algebro-geometric spectral theory
of 2D Schr\"odinger operators was introduced in 1976 by
Dubrovin, Krichever and Novikov~\cite{DKN}. In this theory
periodic 2D Schr\"odinger operators are considered. It turns out
that the Floquet solution of the equation $L\psi=0$ is a Baker-Akhiezer
function on a spectral curve in the space of Floquet multipliers,
and one can reconstruct the operator
starting from its geometric spectral data including the spectral curve, 
the divisor of poles of $\psi$ etc. Later Novikov and Veselov studied 
the case of potential operators~\cite{VN}. 

The inverse spectral problem for the discrete operator~(\ref{discrete}) 
was studied by Krichever~\cite{K}. 
As for the direct spectral problem for the operator~(\ref{discrete}),
it was in fact implicitly studied by one of the authors in
the paper~\cite{O}. Indeed,
the direct spectral problem for 2D discrete elliptic operators with zero 
potential was in fact solved in~\cite{O} by reducing to the direct 
spectral problem
for hyperbolic operators~(\ref{discrete}). We will show how 
to solve the direct spectral problem for operators~(\ref{discrete})
in this paper explicitly.

As far as we know, the algebro-geometric
spectral theory of semi-discrete operators~(\ref{semi-discrete}) was never 
studied. We develop the algebro-geometric spectral theory of 2D semi-discrete
Schr\"odinger operators. The direct spectral
problem is studied using Floquet theory of periodic first order
linear ODEs, and this seems to be new. Despite the fact that an arbitrary
periodic first order linear ODE cannot be solved explicitly, it turns out that
using Floquet theory for linear ODEs 
we can obtain enough spectral information to
understand the structure of the spectral data.

Using these algebro-geometric spectral theories we investigate 
the spectral properties of the Laplace transformations of 
semi-discrete~(\ref{semi-discrete}) 
and discrete~(\ref{discrete}) 
operators. In both cases the Laplace transformations are described as shifts
on the Jacobians of the spectral curves. This makes it possible 
to find solutions of the 
semi-discrete and discrete 2D Toda lattices in terms of theta-functions.
We recall that solutions of the hyperbolic 2D Toda lattice~(\ref{2DToda})
in term of theta-functions were found by 
Krichever~\cite{K1}.

It turns out that in the discrete case
we can give a description of a cyclic chain of Laplace transformations 
in terms of the spectral data and, consequently, in terms of 
the linearizability of the dynamics.

\section{Laplace transformations and 2D Toda lattices}

In this section we start by introducing the Laplace transformations of 
semi-discrete operators~(\ref{semi-discrete}). We show their relation to a 
semi-discrete 2D Toda lattice. Then we recall briefly known results on 
the Laplace transformations of discrete operators~(\ref{discrete}). 

\subsection{Laplace transformations of two-dimensional 
semi-discrete hyperbolic Schr\"odinger operators}

Let us consider an operator of the form~(\ref{semi-discrete}) defined on the
space of (in general complex-valued) functions $\psi_n(y)=\psi(n,y)$ 
defined on $\mathbb{Z}\times\mathbb{R}.$ 
The coefficients $a_n(y),\dots,d_n(y)$ of the operator are also
(in general complex-valued) functions on $\mathbb{Z}\times\mathbb{R}.$ 

Let us define a shift operator $T\psi_n(y)=\psi_{n+1}(y).$

\begin{Lemma}\label{decomposition} 
The operator~(\ref{semi-discrete}) such that $b_n(y)\ne0$ and
$d_n(y)\ne0$ can be uniquely presented in the 
form
$$
L=f_n(y)((\partial_y+A_n(y))(1+v_n(y)T)+w_n(y)).
$$
or in the form
$$
L=\hat{f}_n(y)((1+\hat{v}_n(y)T)(\partial_y +\hat{A}_n(y))+\hat{w}_n(y)),
$$
\end{Lemma}

\noindent{\bf Proof.} Let us write for simplicity $\partial$
instead of $\partial_y$,
$a_n$ instead of $a_n(y)$ etc. We will use $'$ as the derivation
with respect to $y$.
We can obtain by direct calculation that
$$
a_n=f_nA_n+f_nw_n,\quad
b_n=f_n,\quad
c_n=f_nv'_n+f_nA_nv_n,\quad
d_n=f_nv_n.
$$
These equations can be easily solved if $b_n\ne0,$ $d_n\ne0:$ 
$$
f_n=b_n,\quad
v_n=\frac{d_n}{b_n},\quad
A_n=\frac{c_n}{d_n}-\left(\log\frac{d_n}{b_n}\right)',\quad
w_n=\frac{a_n}{b_n}-\frac{c_n}{d_n}+\left(\log\frac{d_n}{b_n}\right)'.
$$
In the same way we can see that if $b_n\ne0,$ $d_n\ne0$ then
$$
\hat{f}_n=b_n,\quad
\hat{v}_n=\frac{d_n}{b_n},\quad
\hat{A}_n=\frac{c_{n-1}}{d_{n-1}},\quad
\hat{w}_n=\frac{a_n}{b_n}-\frac{c_{n-1}}{d_{n-1}}.
$$
This completes the proof. $\Box$

In the following we will consider only operators satisfying the conditions
$b_n\ne0$ and $d_n\ne0.$ 

Let us consider the equation $L\psi=0.$ Using
Lemma~\ref{decomposition} we can define the Laplace transformations.

\begin{Definition} Let us define a Laplace transformation of the
first type as the transformation
$$
L\mapsto\tilde{L}=\tilde{f}_n(w_n(1+v_nT)\frac{1}{w_n}(\partial+A_n)+w_n),\quad
\psi\mapsto\tilde{\psi}=(1+v_nT)\psi,
$$
and a Laplace transformation of the
second type as the transformation
$$
L\mapsto\hat{L}=\hat{f}_n(\hat{w}_n(\partial+\hat{A}_n)\frac{1}{\hat{w}_n}%
(1+\hat{v}_nT)+\hat{w}_n),\quad
\psi\mapsto\hat{\psi}=(\partial +\hat{A}_n)\psi.
$$
Here $\tilde{f}_n(y)$ and $\hat{f}_n(y)$ are arbitrary functions.
\end{Definition}

Since $\tilde{f}_n$ and $\hat{f}_n$ are arbitrary functions,
these transformations define $\tilde{L}$ and $\hat{L}$ up to
a transformation $L\mapsto h_nL.$ But the Laplace transformations
are well-defined transformations of the equation $L\psi=0$.

There are gauge transformations $L\mapsto\bar{L},$ $\psi\mapsto\bar{\psi}$
giving equivalent equations, they are defined by the formulas
$$
\bar{\psi}=g_n^{-1}\psi,\quad
\bar{a}_n=h_na_ng_n+h_nb_ng'_n,\quad
\bar{b}_n=h_nb_ng_n,
$$
$$
\bar{c}_n=h_nc_ng_{n+1}+h_nd_ng'_{n+1},\quad
\bar{d}_n=h_nd_ng_{n+1},
$$
where $h_n(y)\ne0$ and $g_n(y)\ne0$ are arbitrary functions.

Let us remark that a gauge transformation 
$(L,\psi)\mapsto(L',\psi')$ such that $L=L'$ is just a
multiplication of $\psi$ by a constant. For this reason we usually impose
the normalization condition $\psi_0(0)=1.$

As we will see in the next section, for a given periodic operator its
normalized $\psi$-function with prescribed Floquet multipliers is unique.
Hence a pair $(L,\psi)$ is uniquely defined by $L$ in this case, and this case
is the most interesting for us.

For this reason we will view both Laplace transformations and gauge
transformations as transformations of operators $L$ rather than
transformations of pairs $(L,\psi)$ consisting of an operator $L$
and its $\psi$-function.

\begin{Lemma}\label{Laplacegauge}
Laplace transformations of gauge equivalent operators
are gauge equivalent.
\end{Lemma}

\begin{Lemma}\label{canonical}
For each operator there exists a unique gauge equivalent
operator such that 
\begin{equation}\label{condition}
b_n(y)\equiv1,\quad d_n(y)\equiv1.
\end{equation}
\end{Lemma}

\noindent{\bf Proofs} of both Lemmas can be obtained by a direct calculation. 
$\Box$

\begin{Lemma}\label{inverse}
 The Laplace transformations of the first and the second type
are inverse to each other (as transformations of gauge equivalence classes).
\end{Lemma}

\noindent{\bf Proof.} Let us take a gauge equivalence class and take 
the unique operator satisfying the condition~(\ref{condition}) 
in this class as its
representative. For this operator $f_n=v_n=1,$
$A_n=c_n$ and $w_n=a_n-c_n.$ Let us now find its Laplace transformation of the
first type. If we take $\tilde{f}_n=1$ then we obtain an operator
with $\tilde{a}_n=a_n,$ $\tilde{b}_n=1,$ 
$\tilde{c}_n=c_{n+1}\frac{a_n-c_n}{a_{n+1}-c_{n+1}}$
and $\tilde{d}_n=\frac{a_n-c_n}{a_{n+1}-c_{n+1}}.$
For this operator $\hat{f}_n=1,$ $\hat{v}_n=\frac{a_n-c_n}{a_{n+1}-c_{n+1}},$
$\hat{A}_n=c_n$ and $\hat{w}_n=a_n-c_n.$
Let us now find its Laplace transformation of the
second type. If we take $\tilde{f}_n=1$ then we obtain an operator
with $\check{a}_n=a_n-(\log(a_n-c_n))',$ $\check{b}_n=1,$ 
$\check{c}_n=\left(\frac{a_n-c_n}{a_{n+1}-c_{n+1}}\right)'+%
c_{n+1}\frac{a_n-c_n}{a_{n+1}-c_{n+1}}-%
\frac{a_n-c_n}{a_{n+1}-c_{n+1}}(\log(a_n-c_n))',$
and $\check{d}_n=\frac{a_n-c_n}{a_{n+1}-c_{n+1}}.$ The gauge
transformation with $g_n=c_n-a_n$ and $h_n=\frac{1}{c_n-a_n}$ transforms
this operator into the initial one. $\Box$
 
It follows from Lemma~\ref{inverse} that
we can restrict ourselves to the Laplace transformations of the first type.
Given a gauge equivalence class, let us choose $A_n$ 
and $w_n$ of the unique operator
in this class satisfying the condition~(\ref{condition}) as gauge invariants.
They give a complete set of invariants, since 
if the operator satisfies the condition~(\ref{condition}) 
then $f_n=1$ and $v_n=1.$

\begin{Lemma} In terms of the gauge invariants the Laplace transformation
of the first type acts in the following way:
\begin{eqnarray*}
\tilde{A}_n(y)&=&A_{n+1}(y)+\frac{w'_{n+1}(y)}{w_{n+1}(y)},\\
\tilde{w}_n(y)&=&w_n(y)+A_n(y)+\frac{w'_n(y)}{w_n(y)}-%
A_{n+1}(y)-\frac{w'_{n+1}(y)}{w_{n+1}(y)}.
\end{eqnarray*}
\end{Lemma}

\noindent{\bf Proof} can be obtained by direct calculation. $\Box$

Let us now consider a chain $\dots,L_{-1},L_0,L_1,\dots$
of Laplace transformations of the first type,
$L_{k+1}=\tilde{L}_k.$ We obtain the system of equations
\begin{eqnarray*}
A^{k+1}_n&=&A^k_{n+1}+(\log w^k_{n+1})',\\
w^{k+1}_n&=&w^k_n+A^k_n+(\log w^k_n)'-A^k_{n+1}-(\log w^k_{n+1})',
\end{eqnarray*}
where $A^k_n$ and $w_n^k$ are the gauge invariants of $L_k.$
This system is equivalent to the system
\begin{eqnarray}
(\log w^k_{n+1}(y))'&=&A^{k+1}_n(y)-A^k_{n+1}(y),\label{1p}\\
w^{k+1}_n(y)-w^k_n(y)&=&A^{k+1}_{n-1}(y)-A^{k+1}_n(y).\label{2p}
\end{eqnarray}
This system includes only differences of $A^k_n,$ hence we should also fix
some $A^{k_0}_{n_0}$ in order to find all $A^k_n$ from 
equations~(\ref{1p}), (\ref{2p}).
Let us now eliminate $A^k_n.$ Let us remark that given $A^{k+1}_{n-1},$ 
we can find $A^k_{n+1}$ in two different ways: 1) Find $A^{k+1}_n$ 
using~(\ref{2p}), 
then find $A^k_{n+1}$ using~(\ref{1p}), or 2) Find $A^k_n$ using~(\ref{1p}), 
then find $A^k_{n+1}$ using~(\ref{2p}). 
Since the result should be the same, 
this gives us the compatibility condition
\begin{equation}\label{eqw}
w^{k+1}_n-w^k_n-(w^k_{n+1}-w^{k-1}_{n+1})=%
(\log w^k_n)'-(\log w^k_{n+1})'.
\end{equation}
Given a solution of the equation~(\ref{eqw}) and some fixed
$A^{k_0}_{n_0}$, we can find all the
$A^k_n$ using~(\ref{1p}),~(\ref{2p}). 
In this way solving our equations~(\ref{1p}),~(\ref{2p}) reduces to
solving the equation~(\ref{eqw}).

\begin{Theorem} 1) Given a solution of the following 
semi-discrete 2D Toda lattice
\begin{equation}\label{2D1}
(g^k_n-g^k_{n+1})'=e^{g^{k+1}_n-g^k_{n+1}}-e^{g^k_n-g^{k-1}_{n+1}}
\end{equation}
we can obtain a family of chains of Laplace 
transformations of the first type
parameterized by one arbitrary function $A^0_0(y).$

2) Given a chain of Laplace transformations of the first type we can
obtain a family of solutions of the equation~(\ref{2D1}) parameterized
by an arbitrary function $g^0_0(y)$ and a set of 
arbitrary constants $r^k,$ $k\in\mathbb{Z}.$
\end{Theorem}

\noindent{\bf Proof.} If we take the equation~(\ref{2D1}), subtract
the same equation but with the shift $k\mapsto k-1,$ $n\mapsto n+1,$ 
and make the change of variables $w^k_n=e^{g_n^k-g^{k-1}_{n+1}},$ 
then we obtain the equation~(\ref{eqw}). As we explained before,
given a solution of the equation~(\ref{eqw}) and an arbitrary function
$A^0_0(y),$ we can construct a solution of the 
equations~(\ref{1p}), (\ref{2p}) describing a chain of Laplace transformations.
This proves the first statement of the theorem.

Given a solution $w^k_n(y)$ of the equation~(\ref{eqw}), we can 
construct $g^k_n$ starting from $g^0_0(y)$ using
the equations
\begin{equation}\label{t1}
g^k_n(y)-g^{k-1}_{n+1}(y)=\log w^k_n(y),
\end{equation}
\begin{equation}\label{t2}
g^k_n(y)-g^k_{n+1}(y)=\int_0^y(w^{k+1}_n(y')-w^k_n(y'))\,dy'+c^k_n,
\end{equation}
where $c^k_n$ are constants. The solutions $g^k_n$
of the equations~(\ref{t1},\ref{t2}) are clearly solutions of 
the equation~(\ref{2D1}). However, there is a compatibility
condition
$$
c^{k-1}_{n+1}=c^k_n+%
\int_0^y(w^{k+1}_n(y')-w^k_{n+1}(y')-w^k_n(y')+w^{k-1}_{n+1}(y'))\,dy'%
+\log \frac{w^k_{n+1}(y)}{w^k_{n}(y)}
$$
for the equations~(\ref{t1},\ref{t2}). This means that we can take arbitrary
constants $c^k_0=r^k,$ other constants $c^k_n$ are defined by this
compatibility condition. This proves the second statement of the Theorem.
$\Box$

Let us now describe possible modifications in the periodic 
case.
Let us consider a periodic operator with some periods $N$ and $T,$ 
i.e. such that $a_{n+N}(y)=a_n(y),$ 
$a_n(y+T)=a_n(y),\dots$

As was proven
in Lemma~\ref{canonical}, any operator is gauge equivalent to an operator
satisfying the conditions~(\ref{condition}). 
It turns out that this operator is, in general, non-periodic. 

We could consider periodic gauge transformations, i.e. 
such that $h_{n+N}(y)=h_n(y),$ $h_n(y+T)=h_n(y),$
$g_{n+N}(y)=g_n(y),$ $g_n(y+T)=g_n(y),$
in order to preserve the periodicity of operators  
by gauge transformations.
Let us now consider only periodic gauge transformations. 
Lemma~\ref{canonical} does
not hold in this case, it is replaced by the following Lemma.

\begin{Lemma} The function
$I(y)=\frac{b_1(y)\dots b_{N}(y)}{d_1(y)\dots d_{N}(y)}$
is a gauge invariant.
Let $Z(y)$ be a function such that $Z^N(y)=I(y).$ We can transform 
a periodic operator
by a periodic gauge transformation into a unique periodic operator
such that for any $i$ 
\begin{equation}\label{canformper}
b_i(y)=Z(y),\quad d_i(n)=1.
\end{equation}
\end{Lemma}

\noindent{\bf Proof} can be obtained by direct calculation. $\Box$

For an operator satisfying the conditions~(\ref{canformper}) we have
$f_n=Z,$ $v_n=\frac{1}{Z}.$ We take $A_n$ and $w_n$ (as before)
and $I$ as gauge invariants. The choice of $Z(y)$ is not unique,
but $A_n$ and $w_n$ do not depend on this choice. 

We can write now how the Laplace transformation acts in 
terms of $A_n,$ $w_n$ and $Z.$ We
consider only the Laplace transformation of the first type.

\begin{Lemma} The function $I$ is preserved by
the Laplace transformation.
\end{Lemma}
\noindent{\bf Proof} can be obtained by direct calculation. $\Box$

Lemma~\ref{Laplacegauge} holds in the periodic case. Hence we can always
replace an operator by a gauge equivalent one.

Let us take a periodic operator. Let us choose $Z$ such that
$Z^N=I$ and transform our operator to the operator satisfying the 
conditions~(\ref{canformper}). Let us apply to this operator the
Laplace transformation. The invariant $I$ is preserved. We transform the
resulting operator
by a gauge transformation into the operator satisfying~(\ref{canformper}) 
with the same $Z.$ This gives us a transformation of $A_n$ and $w_n.$
We obtain the following Lemma

\begin{Lemma} The Laplace transformation acts in the following way:
\begin{eqnarray*}
\tilde{A}_n(y)&=&A_{n+1}(y)+\frac{w'_{n+1}(y)}{w_{n+1}(y)}+(\log Z(y))',\\
\tilde{w}_n(y)&=&w_n(y)+A_n(y)%
+\frac{w'_n(y)}{w_n(y)}-A_{n+1}(y)-\frac{w'_{n+1}(y)}{w_{n+1}(y)}-(\log Z(y))'.
\end{eqnarray*}
The compatibility condition for these equations is given by the 
equation~(\ref{eqw}).
\end{Lemma}

\noindent{\bf Proof} can be obtained by a direct calculation. 
$\Box$

This means that in the periodic case we obtain the same equation~(\ref{eqw})
describing the compatibility condition, as in the general case.
The only difference is that 
$w^k_n(y)$ should be periodic in $n$ and $y.$

\subsection{Laplace transformations of two-dimensional discrete hyperbolic
Schr\"odinger operators}\label{discretelaplace}

We recall here briefly some already known results
following the paper~\cite{ND}, where one 
can find proofs and a more extended exposition.
Let us consider the shift operators
$$
T_1\psi(n,m)=\psi(n+1,m),\quad%
T_2\psi(n,m)=\psi(n,m+1)
$$
acting on functions defined on $\mathbb{Z}^2.$ We can rewrite the 
operator~(\ref{discrete}) as
\begin{equation}\label{discrete1}
L\psi=(a_{n,m}+b_{n,m}T_1+c_{n,m}T_2+%
d_{n,m}T_1T_2)\psi.
\end{equation}
We are interested in the equation 
\begin{equation}\label{discreteequation}
L\psi=(a_{n,m}+b_{n,m}T_1+c_{n,m}T_2+%
d_{n,m}T_1T_2)\psi=0.
\end{equation}
There are gauge transformations
\begin{equation}\label{gaugediscrete}
L\mapsto f_{n,m}Lg_{n,m},\quad \psi_{n,m}=g^{-1}_{n,m}\psi_{n,m}
\end{equation}
giving equivalent equations.

The operator~(\ref{discrete1}) can be presented uniquely in the form
\begin{equation}\label{repr12}
L=f_{n,m}((1+u_{n,m}T_1)(1+v_{n,m}T_2)+w_{n,m}).
\end{equation}
The operator~(\ref{discrete1}) can also be presented uniquely in the form
\begin{equation}\label{repr21}
L=f'_{n,m}((1+v'_{n,m}T_2)(1+u'_{n,m}T_1)+w'_{n,m}).
\end{equation}
We can define a Laplace transformation of the first type
$$
L\mapsto \tilde{L}=\tilde{f}_{n,m}(w_{n,m}(1+v_{n,m}T_2)w^{-1}_{n,m}%
(1+u_{n,m}T_1)+w_{n,m}),
$$
$$
\psi\mapsto \tilde{\psi}=(1+v_{n,m}T_2)\psi
$$
and we can define a Laplace transformation of the second type
$$
L\mapsto L'=f'_{n,m}(w'_{n,m}(1+u'_{n,m}T_1)(w'_{n,m})^{-1}%
(1+v'_{n,m}T_2)+w'_{n,m}),
$$
$$
\psi\mapsto\psi'=(1+u'_{n,m}T_1)\psi.
$$
These transformations transform gauge equivalent operators
into gauge equivalent operators, i.e. they act on gauge equivalence classes.
The transformations of the first and the second type are inverse to each other
(as transformations of gauge equivalence classes). It follows that we can
restrict ourselves to the Laplace transformations of the first type.

Let us introduce gauge invariants in the following manner. We can transform
an operator by a gauge transformation to an operator such that $f_{n,m}=1.$
Then we can take $w_{n,m}$ and $H_{n,m}=\frac{v_{n,m}u_{n,m+1}}{u_{n,m}v_{n+1,m}}$
of this operator as gauge invariants.

In terms of the gauge invariants the Laplace 
transformation of the first type has the form
\begin{equation}\label{tr1}
1+\tilde{w}_{n+1,m}=(1+w_{n,m+1})%
\frac{w_{n,m}w_{n+1,m+1}}{w_{n+1,m}w_{n,m+1}}H^{-1}_{n,m},
\end{equation}
\begin{equation}\label{tr2}
\tilde{H}_{n,m}=\frac{1+w_{n,m+1}}{1+\tilde{w}_{n,m+1}}.
\end{equation}
Let us now consider a chain of Laplace transformations of the first type;
we obtain $H^{(k+1)}_{n,m}=\tilde{H}^{(k)}_{n,m},$ 
$w^{(k+1)}_{n,m}=\tilde{w}^{(k)}_{n,m}.$ After excluding $H^{(k)}_{n,m}$
from equations~(\ref{tr1}), (\ref{tr2}) we obtain 
the so-called completely discretized 2D Toda lattice
\begin{equation}\label{completelydiscr}
\frac{1+w^{(k+2)}_{n+1,m}}{1+w^{(k+1)}_{n+1,m}}%
\frac{1+w^{(k+1)}_{n,m+1}}{1+w^{(k)}_{n,m+1}}=%
\frac{w^{(k)}_{n+1,m}w^{(k)}_{n,m+1}}{w^{(k)}_{n,m}w^{(k)}_{n+1,m+1}}.
\end{equation}

We defined the Laplace transformation of the first type using 
the representation~(\ref{repr12}) where we used the shift $T_1$ and 
then the shift $T_2.$ Let us denote this transformation
by $\Lambda^{++}_{12}.$ The Laplace transformation of the second type
was defined using the representation~(\ref{repr21}) where we used 
the shift $T_2$ and then the shift $T_1.$ Let us denote 
this transformation by $\Lambda^{++}_{21}.$ We can however take any pair
of orthogonal shifts $T^\pm_i,$ $T^\pm_j,$ $i\ne j.$
Hence we can introduce in an analogous way Laplace transformations
corresponding to any pair of orthogonal shifts
$$
\begin{array}{rr}
(T_1,T_2),(T_2,T_1)\rightarrow\Lambda_{12}^{++}, \Lambda_{21}^{++};&
(T_1^{-1},T_2),(T_2,T_1^{-1})\rightarrow \Lambda_{12}^{-+}, \Lambda_{21}^{+-};\\
(T_1,T_2^{-1}),(T_2^{-1},T_1)\rightarrow \Lambda_{12}^{+-}, \Lambda_{21}^{-+};&
(T_1^{-1},T_2^{-1}),(T_2^{-1},T_1^{-1})\rightarrow \Lambda_{12}^{--},
\Lambda_{21}^{--}.
\end{array}
$$
It is easy to see that $\Lambda_{12}^{st}\Lambda_{21}^{ts}=1$ for $s,t=\pm$.

Let us also introduce a transformation $S_1$:
$$
L\mapsto\tilde L=\tilde
f_{n,m}(a_{n-1,m}+b_{n-1,m}T_1+c_{n-1,m}T_2+d_{n-1,m}T_1T_2),
$$
$$
\psi_{n,m}\mapsto\tilde\psi_{n,m}=\psi_{n-1,m}
$$
and a transformation $S_2$:
$$
L\mapsto\tilde L=\tilde
f_{n,m}(a_{n,m-1}+b_{n,m-1}T_1+c_{n,m-1}T_2+d_{n,m-1}T_1T_2),
$$
$$
\psi_{n,m}\mapsto\tilde\psi_{n,m}=\psi_{n,m-1}.
$$

It is clear that $S_k$ commutes with  $\Lambda_{ij}^{st}$. 
As usual, we consider all transformations as transformations of 
gauge equivalence classes. 
A direct calculation leads us to
the following Lemma.
\begin{Lemma}\label{group} The following identities hold:
\begin{align*}
\Lambda_{12}^{++}&=S_1\Lambda_{12}^{-+}&
\Lambda_{12}^{++}&=S_2\Lambda_{12}^{+-}&
\Lambda_{12}^{++}&=S_2S_1\Lambda_{12}^{--}
\end{align*}
\end{Lemma}

This means that the group of transformations generated by
$\Lambda^{st}_{ij}$ has three 
generators.

\section{Algebro-geometric spectral theory of
two-dimensional semi-discrete and discrete hyperbolic Schr\"odinger operators}

In this section we start by an investigation of the
algebro-geometric spectral theory
of semi-discrete operators~(\ref{semi-discrete}). Then we consider
an algebro-geometric theory of discrete operators~(\ref{discrete}).
We recall known results on the inverse
spectral problem for discrete operators~(\ref{discrete}) due to 
Krichever~\cite{K}. We investigate the direct spectral problem
for discrete operators. This was already done implicitly by 
one of the authors in the paper~\cite{O}. 

\subsection{Algebro-geometric spectral theory of two-dimensional 
semi-discrete hyperbolic Schr\"odinger operators}

Let us consider operators $L$ of the form~(\ref{semi-discrete}) and
the corresponding equations $L\psi=0.$

\subsubsection{Direct spectral problem} Let us consider
a periodic operator $L,$ i.e. such that the functions $a_n(y),\dots,d_n(y)$
satisfy the conditions $a_{n+N}(y)=a_n(y),$ $a_n(y+T)=a_n(y),\dots.$
We will also consider only periodic gauge transformations.

We will consider only periodic operators 
such that
the gauge invariant $I(y)$ is a constant. As any operator is gauge equivalent
to the operator satisfying the 
conditions~(\ref{canformper}), we will consider only 
the operators satisfying these conditions. Operators with different $I$
are not equivalent, but their spectral theory is the same. 
We will consider for
simplicity the case of $Z=-1,$ i.e. we consider periodic operators
of the form
\begin{equation}\label{oursemi-discrete}
(L\psi)_n=a_n(y)\psi_n(y)-\psi'_n(y)+c_n(y)\psi_{n+1}(y)+\psi'_{n+1}(y)
\end{equation}
and the corresponding equations $L\psi=0.$

\begin{Definition} Let $\rho$ and $\mu$ be two 
complex numbers. A solution $\psi_n(y)$ of the equation $L\psi=0$ 
is said to be a Floquet solution with Floquet multipliers $\rho$ with
respect to $n$ and $\mu$ with respect to $y,$ if for any $n$ and $y$ we have 
\begin{equation}\label{Floquet}
\psi_{n+N}(y)=\rho\psi_n(y),\quad\psi_n(y+T)=\mu\psi_n(y).
\end{equation}
\end{Definition}

Since $\psi_n(y)$ is defined up to multiplication by a constant, we will
impose the normalization condition $\psi_0(0)=1.$

Our first goal is to describe possible pairs $(\rho,\mu)$ of
Floquet multipliers. 
Let us fix some Floquet multiplier $\rho.$ 
It follows from~(\ref{Floquet}) that any $\psi_n(y)$ 
can be expressed using $\rho$
and $\psi_0(y),\dots,\psi_{N-1}(y).$ Thus the equation $L\psi=0$
is equivalent to a finite number of linear 
ODEs on $\psi_0(y),\dots,\psi_{N-1}(y).$ It is easy to see that $L\psi=0$
is equivalent to the equation  $B\Psi'(y)+C(y,\rho)\Psi(y)=0,$ where
$$
\Psi(y)=\left(%
\begin{array}{c}
\psi_0(y)\\
\dots\\
\psi_{N-1}(y)
\end{array}\right),\quad
B=\left(%
\begin{array}{cccccc}
-1&1&0&\dots&0&0\\
0&-1&1&\dots&0&0\\
\dots&\dots&\dots&\dots&\dots&\dots\\
0&0&0&\dots&-1&1\\
\rho&0&0&\dots&0&-1
\end{array}%
\right),
$$
$$
C(y,\rho)=\left(%
\begin{array}{cccccc}
a_0&c_0&0&\dots&0&0\\
0&a_1&c_1&\dots&0&0\\
\dots&\dots&\dots&\dots&\dots&\dots\\
0&0&0&\dots&a_{N-2}&c_{N-2}\\
\rho c_{N-1}&0&0&\dots&0&a_{N-1}
\end{array}%
\right).
$$
It follows that the equation $L\psi=0,$ where $\psi$ has a
Floquet multiplier $\rho,$ is equivalent to the
linear ODE 
\begin{equation}\label{ourODE}
\Psi'(y)=A(y,\rho)\Psi(y),
\end{equation}
where $A(y,\rho)=-B^{-1}(\rho)C(y).$
The Floquet multiplier $\rho$ enters in this linear ODE as a parameter.
It is easy to check that 
$$
B^{-1}=\frac{1}{\rho-1}\left(%
\begin{array}{ccccc}
1&1&\dots&1&1\\
\rho&1&\dots&1&1\\
\dots&\dots&\dots&\dots&\dots\\
\rho&\rho&\dots&\rho&1
\end{array}%
\right).
$$
Thus for $\rho\ne1$ the function $A(y,\rho)$ is holomorphic with respect to
$\rho.$ Let us remark that $A(y,\rho)$ is periodic: $A(y+T,\rho)=A(y,\rho).$
We will occasionally omit $y$ to shorten the notation.

\begin{Definition} Let $\mu$ be a complex number.
A solution $\Psi(y)$ of the equation $\Psi'=A\Psi$ 
is said to be a Floquet solution with Floquet multiplier
$\mu$ if for any $y$ we have 
\begin{equation}\label{Floquet2}
\Psi(y+T)=\mu\Psi(y).
\end{equation}
\end{Definition}

We see that the question of describing possible 
Floquet multipliers $\rho\ne1,$ $\mu$ for the equation $L\psi=0$
can be restated in the following way:
given $\rho\ne1,$ for which $\mu$ does there exist a Floquet solution of the
periodic equation $\Psi'=A(\rho)\Psi?$ This permits us to use
the Floquet theory of periodic linear differential equations.

Let us recall some standard facts from this theory, see e.g.~\cite{R}. 
Let us consider a homogeneous linear ODE
\begin{equation}\label{linearhom}
\frac{dx(t)}{dt}=A(t)x(t),
\end{equation}
where $x\in\mathbb{R}^n$ and $A(t)$ is an $n\times n$-matrix. 
An $n\times n$-matrix $\Phi(t,s)$ is called a resolvent of $A(t)$
if $\phi(t)=\Phi(t,t_0)x_0$ is the solution of~(\ref{linearhom})
satisfying the initial condition $\phi(t_0)=x_0.$ The resolvent
exists and is uniquely determined by the following properties:
$$
\forall t\,\,\,\Phi(t,t)=I,\quad%
\forall s,t,u\,\,\,\Phi(t,s)=\Phi(t,u)\Phi(u,s),
$$
\begin{equation}\label{rez}
\frac{\partial}{\partial t}\Phi(t,s)=A(t)\Phi(t,s),\quad%
\frac{\partial}{\partial s}\Phi(t,s)=-\Phi(t,s)A(s).
\end{equation}
If $\hat{\Phi}(t)$ is a fundamental matrix of 
the equation~(\ref{linearhom}) 
(i.e. a matrix such that
its columns form a basis in the space of solutions of this
equation) then
\begin{equation}\label{fund}
\Phi(t,s)=\hat{\Phi}(t)\hat{\Phi}^{-1}(s).
\end{equation}

Let us now consider an inhomogeneous linear ODE
\begin{equation}\label{lininh}
\frac{dx(t)}{dt}=A(t)x(t)+b(t).
\end{equation}
The solution $\phi(t)$ of this equation satisfying
the initial condition $\phi(t_0)=x_0,$ 
can be found with the help of the resolvent of $A(t):$
\begin{equation}\label{solinh}
\phi(t)=\Phi(t,t_0)x_0+\int_{t_0}^t\Phi(t,s)b(s)\,ds.
\end{equation}

Let the matrix $A(t)$ be periodic: $A(t+T)=A(t).$
It follows that for any $n\in\mathbb{N},$ $t$ we have
$\Phi(t+nT,0)=\Phi(t,0)\Phi(nT,0).$ This implies
the following Lemma.

\begin{Lemma}\cite{R}\label{Floquetsum}
A solution $g(t)$ of the periodic
equation~(\ref{linearhom}) is a Floquet solution with
a Floquet multiplier $\mu$
$$
\forall t\,\,\,g(t+T)=\mu g(t)
$$
if and only if the initial condition $g_0=g(0)$
is an eigenvector of the matrix $\Phi(T,0)$
with the eigenvalue $\mu:$
$$
\Phi(T,0)g_0=\mu g_0.
$$
\end{Lemma}

Let $L$ be a generic operator.
Consider the equation~(\ref{ourODE}). It follows from
Lemma~\ref{Floquetsum} that given $\rho\ne1,$ we have
$N$ Floquet multipliers $\mu_1,\dots,\mu_N$ (possibly coinciding)
corresponding to independent Floquet solutions. Let us now recall
that for $\rho\ne1$ the matrix $A(y,\rho)$ is holomorphic
with respect to $\rho.$ It follows
that solutions of the equation~(\ref{ourODE}) are also holomorphic
functions of $\rho.$ The formula~(\ref{fund})
expressing $\Phi(t,s)$ in terms
of a fundamental matrix implies that the resolvent $\Phi(t,s,\rho)$
of $A(y,\rho)$ is also a holomorphic function of $\rho.$ We obtain the
following Lemma.

\begin{Lemma} Let $L$ be a generic periodic operator.
The possible pairs $(\rho,\mu)$ of Floquet multipliers of the
equation $L\psi=0,$ such that $\rho\ne1,$ form an analytic
curve $\tilde{\Gamma},$ called a spectral curve. This curve is
given by the equation $\det(\Phi(T,0,\rho)-\mu I)=0.$
The natural projection $\pi:(\rho,\mu)\mapsto\rho$ gives us
an $N$-fold covering 
$\pi:\tilde{\Gamma}\longrightarrow\mathbb{C}\setminus \{1\}.$
\end{Lemma}

Let us now consider the eigenvectors $\Psi(0)$ of 
the matrix $\Phi(T,0,\rho).$
Since the matrix $\Phi(T,0,\rho)$ is holomorphic 
with respect to $\rho,$
its eigenvectors $\Psi(0)$ (let us recall that we imposed the
condition $(\Psi(0))_0=\psi_0(0)=1$)
are meromorphic functions
 on the spectral curve $\tilde{\Gamma}.$

\begin{Lemma}\label{psi-function}
The solution $\psi_n(y)$ of the equation $L\psi=0$
is a meromorphic function on the spectral curve $\tilde{\Gamma}.$ 
Its poles does not depend on $y.$
\end{Lemma}

\noindent{\bf Proof} follows from the fact that 
$\Psi(y)=\Phi(y,0,\rho)\Psi(0)$ and $\Phi(y,0,\rho)$ is
a holomorphic function in $\rho.$ $\Box$

We will consider $\psi_n(y)$ as a function defined on the spectral curve
and depending on the parameters $n$ and $y.$ Sometimes we will write
explicitly $\psi_n(y,P),$ where $P$ is a point on the spectral curve.

Let us consider the simplest case when the coefficients $a_n(y),$
$c_n(y)$ of the operator~(\ref{ourODE}) do not depend on $y.$
In this case the matrix $A(y,\rho)$ does not depend on $y$
and we can easily solve the equation~(\ref{ourODE}): 
$\Psi(y)=e^{A(\rho)T}\Psi(0).$ It follows that
the equation of the spectral curve $\tilde{\Gamma}$ is
$\det(e^{A(\rho)T}-\mu I)=0.$ In this case it is better to use other 
coordinates, one can consider a curve $\det(A(\rho)-zI)=0$ and consider
$\mu=e^{zT}$ as a function on this curve.

In the general case we cannot solve the equation~(\ref{ourODE}) and find
the spectral curve explicitly. However, we can obtain enough
information about the spectral curve and the solution $\psi_n(y)$
in order to understand what kind of spectral data we should consider
in the inverse spectral problem.

For negative $n$ we define 
a pole of order $n$ as a zero of order $-n$
and a zero of order $n$ as a pole of order $-n.$

Let us make the following observation: if $\rho=0$ then 
the matrix $A(y,\rho)$ is an upper-triangular matrix. This permits us
to prove the following Lemma

\begin{Lemma}\label{Pplus}
 The fiber $\pi^{-1}(0)$ of the projection
$\pi:\tilde{\Gamma}\longrightarrow\mathbb{C}\setminus \{1\}.$
consists of points $P^+_i=(0,e^{\int_0^Ta_{i-1}(\xi)\,d\xi}),$
$i=1,\dots,N.$ For a generic operator $L$ 
the function $\psi_n(y)$ has a zero of
order $\left[\frac{n-i}{N}\right]+1$ at the point $P^+_i.$
\end{Lemma}

If there are coinciding points $P^+_i$ we should treat these points
with multiplicities, i.e. we should add the corresponding orders of zero. 

\noindent{\bf Proof.} We will consider for simplicity the case $N=2.$
The proof is analogous for $N>2.$

We have 
$$
A(y,0)=\left(%
\begin{array}{cc}
a_0(y)&a_1(y)+c_0(y)\\
0&a_1(y)
\end{array}%
\right).
$$

Let us put $\alpha={a_0},$ $\beta=a_1+c_0,$
$\gamma=a_1$ in order to shorten the notation.
Then our equation~(\ref{ourODE}) for $\rho=0$ becomes the equation
\begin{equation}\label{floquet0}
\left(%
\begin{array}{c}
\psi_0\\
\psi_1
\end{array}\right)'=\left(%
\begin{array}{cc}
\alpha&\beta\\
0&\gamma
\end{array}%
\right)%
\left(%
\begin{array}{c}
\psi_0\\
\psi_1
\end{array}\right),
\end{equation}
The second equation $\psi'_1(y)=\gamma(y)\psi_1(y)$
is easy to solve: $\psi_1(y)=e^{\int_{y_0}^y\gamma(\xi)\,d\xi}\psi_1(y_0).$
This means that $\tilde{\Phi}(t,s)=e^{\int_{s}^t\gamma(\xi)\,d\xi}$
is a resolvent of $\gamma(y).$
Let us now consider the first equation in~(\ref{floquet0}):
\begin{equation}\label{eq1}
\psi'_0(y)=\alpha(y)\psi_0(y)+\beta(y)\psi_1(y).
\end{equation}
The resolvent of $\alpha(y)$ is 
$\hat{\Phi}(t,s)=e^{\int_{s}^t\alpha(\xi)\,d\xi}.$
Using $\hat{\Phi}(t,s)$ and the formula~(\ref{solinh}) 
we can solve the equation~(\ref{eq1}):
$$
\psi_0(y)=e^{\int_{y_0}^y\alpha(\xi)\,d\xi}\psi_0(y_0)+%
\int_{y_0}^ye^{\int_\eta^y\alpha(\xi)\,d\xi}\beta(\eta)\psi_1(\eta)\,d\eta=
$$
$$
=e^{\int_{y_0}^y\alpha(\xi)\,d\xi}\psi_0(y_0)+%
\int_{y_0}^ye^{\int_\eta^y\alpha(\xi)\,d\xi+\int_{y_0}^\eta\gamma(\xi)\,d\xi}%
\beta(\eta)\,d\eta\,\psi_1(y_0).
$$

These explicit formulas for $\psi_i(y)$ at $\rho=0$ 
give immediately an explicit formula for $\Phi(T,0,0):$
\begin{equation}\label{Phi}
\Phi(T,0,0)=\left(%
\begin{array}{cc}
e^{\int_{0}^T\alpha(\xi)\,d\xi}&%
\int_{0}^Te^{\int_\eta^T\alpha(\xi)\,d\xi+\int_{0}^\eta\gamma(\xi)\,d\xi}%
\beta(\eta)\,d\eta\\
0&
e^{\int_{0}^T\gamma(\xi)\,d\xi}
\end{array}%
\right).
\end{equation}

This formula gives us explicitly that the fiber
$\pi^{-1}(0)$ of the projection
$\pi:\tilde{\Gamma}\mapsto\mathbb{C}\setminus \{1\}$
consists of points $P^+_1=(0,e^{\int_0^Ta_0(\xi)\,d\xi}),$
and $P^+_2=(0,e^{\int_0^Ta_1(\xi)\,d\xi}).$

To shorten the notation, let us denote
the matrix elements of $\Phi(T,0,0)$ by $\mu_1,$
$\mu_2$ and $\delta,$ i.e. 
$$
\Phi(T,0,0)=\left(%
\begin{array}{cc}
\mu_1&\delta\\
0&\mu_2
\end{array}%
\right).
$$

Let us consider the generic case where
$\mu_1\ne\mu_2$ and $\delta\ne0.$
The matrix $\Phi(T,0,0)$ has two eigenvectors (normalized as usual by
the condition $\psi_0(0)=1$)
$$
\left(%
\begin{array}{c}
1\\
0
\end{array}%
\right)\quad\mbox{and}\quad
\left(%
\begin{array}{c}
1\\
\frac{\mu_2-\mu_1}{\delta}
\end{array}%
\right)
$$
with the eigenvalues $\mu_1$ and $\mu_2$
respectively.

Let us now consider a neighborhood of $\rho=0.$ Since $\Phi(T,0,\rho)$ is
holomorphic and hence can be expanded in a power series 
in $\rho,$ the normalized eigenvectors
can also be expanded in a series in powers of $\rho:$
$$
\left(%
\begin{array}{c}
1\\
C_1\rho+\dots
\end{array}%
\right)\quad\mbox{and}\quad
\left(%
\begin{array}{c}
1\\
\frac{\mu_2-\mu_1}{\delta}+C_2\rho+\dots
\end{array}%
\right),
$$
where $C_1$ and $C_2$ are constants.

In the generic case $\mu_1\ne\mu_2,$ i.e. $\rho=0$ is not a
branching point. It follows that $\rho$ is a local parameter in a
neighborhood of $P^+_1=(0,\mu_1)$ and $P^+_2=(0,\mu_2).$

We obtain that in a neighborhood of $P^+_1$
$$
\psi_0(0)=1,\quad
\psi_1(0)=C_1\rho+\dots,
$$
$$
\psi_2(0)=\rho\Psi_0(0)=\rho,\quad
\psi_3(0)=\rho\psi_1(0)=C_1\rho^2+\dots,\quad\dots
$$
i.e. $\psi_0(0)$ has a zero of order $0$ in $P^+_1,$
$\psi_1(0)$ has a zero of order $1$ in $P^+_1,$
$\psi_2(0)$ has a zero of order $1$ in $P^+_1,$
$\psi_3(0)$ has a zero of order $2$ in $P^+_1,\dots$

Let us now recall that $\Phi(y,0,0)$ is an upper-triangular matrix.
It follows that
$$
\Phi(y,0,\rho)=\left(%
\begin{array}{cc}
C_1+\dots & C_2+\dots\\
\rho C_3+\dots & C_4+\dots,
\end{array}%
\right),
$$
where $C_1,\dots,C_4$ do not depend on $\rho.$
Then the function
$\psi_0(y)=C_1\psi_0(0)+C_2\psi_1(0)=C_1+\dots$ has a zero of the
same order in $P^+_1,$ as the function $\psi_0(0).$ 
The same argument works for any $\psi_i(y).$ 
We see that the orders of zeroes of $\psi_i(y)$ in
$P^+_1$ are as described in the statement of this Lemma. 
The proof for $P^+_2$ is analogous. $\Box$

In general the number of branching points of the covering
$\pi:\tilde{\Gamma}\longrightarrow\mathbb{C}\setminus \{1\}$
is infinite. We will say that an operator is algebro-geometric
if this covering has only a finite number of branching points.
Let us now consider only this case. We will show that in this case we can 
compactify $\tilde{\Gamma}$ and obtain a compact Riemann surface $\Gamma$
and a covering $\pi:\Gamma\longrightarrow\mathbb{C}P^1.$

\begin{Lemma}\label{Pminus}
Let $L$ be a generic algebro-geometric operator. Then we can
add to $\tilde{\Gamma}$ points 
$P^-_i=(\infty,e^{-\int_0^Tc_{N-i}(\xi)\,d\xi}),$ 
$i=1,\dots,N,$ in such a way that we obtain an analytic curve $\hat{\Gamma}$
with a projection $\pi:\hat{\Gamma}\longrightarrow\mathbb{C}P^1\setminus\{1\}.$
The function $\psi_n(y)$ can be continued
to all of $\hat{\Gamma}.$ 
The function $\psi_n(y)$ has a pole of order 
$\left[\frac{n-i}{N}\right]+1$ at the point $P^-_i.$
\end{Lemma}

\noindent{\bf Proof.} Let us compactify the $\rho$-plane $\mathbb{C}$
by adding the point $\infty$ at infinity with local parameter $t=\frac{1}{\rho}.$
It is easy to check that the matrix $A(y,\rho)$ has a limit when
$\rho$ tends to infinity. We can analytically continue
$A(y,\rho)$ to a holomorphic function on $\mathbb{C}P^1\setminus\{1\}.$
Moreover, the matrix $A(y,\infty)$ is a lower-triangular matrix. We can
prove in the same way as in Lemma~\ref{Pplus} that the 
Floquet multipliers at $\infty$ are $e^{-\int_0^Tc_{N-i}(\xi)\,d\xi},$
$i=1,\dots,N.$ Since our operator is algebro-geometric, there is
a neighborhood of $\infty$ without branching points. Using this 
fact we can add
points $P^-_i$ in order to obtain an analytic curve $\hat{\Gamma}.$ If
all Floquet multipliers are different, we add these points with the
local parameter $t.$ If there are coinciding $P^-_i,$ we add them with the 
local parameter equal to the root of $t$ of degree corresponding to 
the multiplicity of these Floquet multipliers. The calculation of the
order of poles of $\psi_n(y)$ in these points can be done in the
same way as in Lemma~\ref{Pplus}. $\Box$ 

\begin{Lemma}\label{pointQ}
Let $L$ be a generic algebro-geometric operator. 
Then we can
add the point $(1,1)$ and a point $Q$ to $\hat{\Gamma}$ 
in such a way that we obtain an analytic compact 
Riemann surface $\Gamma$
with a projection $\pi:\Gamma\longrightarrow\mathbb{C}P^1.$
The function $\psi_n(y)$ can be continued
to all of $\Gamma.$ 
The function $\psi_n(y)$ is a Baker-Akhiezer function on $\Gamma,$
i.e. it is meromorphic on $\Gamma\setminus Q$ and has
an exponential singularity at the point $Q.$
The function $\psi_n(y)$ can be presented 
in a neighborhood of $Q$ as $e^\frac{Ky}{t}h_n(y,t),$
where $K$ is a constant, $t$ is a local parameter at $Q$
and $h_n(y,t)$ is a function holomorphic with respect to $t,$ 
such that $h(y,0)\ne0.$
The Floquet multiplier $\mu$ in a neighborhood of $Q$ can
be presented as $e^\frac{M}{t}g(t),$ where $M$ is a
constant and $g(t)$ is a
holomorphic function, such that $g(0)\ne0.$
\end{Lemma}

\noindent{\bf Proof.} We will consider for simplicity the case $N=3.$
The proof is analogous for $N\ne3.$

The matrix $A(y,\rho)$ has a pole at $\rho=1.$
Let  $t=\rho-1,$ then 
$$
A(y,t)=\frac{1}{t}A_{-1}(y)+A_0(y),
$$ 
where
$$
A_{-1}(y)=-\left(%
\begin{array}{ccc}
a_0(y)+c_2(y) & a_1(y)+c_0(y) & a_2(y)+c_1(y)\\
a_0(y)+c_2(y) & a_1(y)+c_0(y) & a_2(y)+c_1(y)\\
a_0(y)+c_2(y) & a_1(y)+c_0(y) & a_2(y)+c_1(y)
\end{array}%
\right).
$$
The matrix $A(y,t)$ tends to $\frac{1}{t}A_{-1}(y)$
as $t$ tends to zero, thus
the solution of the equation~(\ref{ourODE}) tends to 
the solution of the equation
\begin{equation}\label{ourODEQ}
\Psi'(y)=\frac{1}{t}A_{-1}(y)\Psi.
\end{equation}
The key observation is that the matrix $A_{-1}(y)$ has the same
rows. It follows that $\psi'_0(y)-\psi'_1(y)=0,$  $\psi'_0(y)-\psi'_2(y)=0.$
This implies that $\psi'_0(y)-\psi'_1(y)$ and $\psi'_0(y)-\psi'_2(y)$
are constants that can be expressed in terms of the initial conditions.
It follows that
$$
\psi_1(y)=\psi_0(y)-\psi_0(0)+\psi_1(0),\quad
\psi_2(y)=\psi_0(y)-\psi_0(0)+\psi_2(0).
$$
We substitute these formulas in the equation for $\psi'_0(y).$
Let us put
$$
E(y)=-a_0(y)-c_2(y),\quad F(y)=-a_1(y)-c_0(y),\quad G(y)=-a_2(y)-c_1(y)
$$
in order to shorten the notation.
We obtain the equation
$$
\textstyle
\psi'_0(y)=\frac{E(y)+F(y)+G(y)}{t}%
\psi_0(y)-\frac{F(y)+G(y)}{t}\psi_0(0)%
+\frac{F(y)}{t}\psi_1(0)+\frac{G(y)}{t}\psi_2(0).
$$
This is a first-order linear ODE which can be explicitly solved.
Using its solution we can find $\psi_i(y)$ and find explicitly
the resolvent $\Phi(y,0).$
Let us introduce the notation
$$
P(y)=e^{\int_0^y\frac{E(\xi)+F(\xi)+G(\xi)}{t}\,d\xi}+%
\int_0^ye^{\int_\eta^y\frac{E(\xi)+F(\xi)+G(\xi)}{t}\,d\xi}%
\frac{-F(\eta)-G(\eta)}{t}\,d\eta,
$$
$$
Q(y)=\int_0^ye^{\int_\eta^y\frac{E(\xi)+F(\xi)+G(\xi)}{t}\,d\xi}%
\frac{F(\eta)}{t}\,d\eta,
$$
$$
R(y)=\int_0^ye^{\int_\eta^y\frac{E(\xi)+F(\xi)+G(\xi)}{t}\,d\xi}%
\frac{G(\eta)}{t}\,d\eta.
$$
We obtain 
$$
\Phi(T,0)=\left(%
\begin{array}{ccc}
P(T) & Q(T) & R(T)\\
P(T)-1 & Q(T)+1 & R(T)\\
P(T)-1 & Q(T) & R(T)+1
\end{array}%
\right).
$$
This matrix has eigenvalues $\mu_1=1,$ $\mu_2=1$ and $\mu_3=P(T)+Q(T)+R(T).$
We see that $\mu_3$ tends to the infinity as $t$ tends to zero. 
We can add to the curve $\hat{\Gamma}$ the point $(1,1)$ with 
a local parameter
$\sqrt{t}$ on the branches
where $\mu$ tends to $1$ and a point $Q$ with a local parameter $t$
on the branch where $\mu$ tends to the infinity. We obtain an
analytic compact
Riemann surface $\Gamma$ and
a holomorphic projection $\pi:\Gamma\longrightarrow\mathbb{C}P^1.$
One can easily verify that $\psi_n(y)$ can be continued as a meromorphic
function to the point $(1,1).$

Let us now consider the point $Q.$ The eigenvector corresponding
to $\mu_3$ is
$\left(%
\begin{array}{ccc}
1&1&1
\end{array}%
\right)^T.$
Using formulas for $P,$ $Q$ and $R$ we see that for the solution
of the equation~(\ref{ourODEQ}) we have the formula
$$
\left(%
\begin{array}{c}
\psi_0(y)\\
\psi_1(y)\\
\psi_2(y)
\end{array}%
\right)=\left(%
\begin{array}{ccc}
P(y) & Q(y) & R(y)\\
P(y)+1 & Q(y)+1 & R(y)\\
P(y)-1 & Q(y) & R(y)+1
\end{array}%
\right)\left(%
\begin{array}{c}
1\\
1\\
1
\end{array}%
\right)=
$$
$$
=e^{\int_0^y\frac{E(\xi)+F(\xi)+G(\xi)}{t}\,d\xi}\left(%
\begin{array}{c}
1\\
1\\
1
\end{array}%
\right).
$$
We see that the solution
$\psi_n(y)$ of the equation~(\ref{ourODE}) in a neighborhood of 
$Q$ has the same behavior as 
$e^{\int_0^y\frac{E(\xi)+F(\xi)+G(\xi)}{t}\,d\xi}$ as $t$ tends to zero.
It is easy to see that $E+F+G=-(c_0+c_1+c_2-a_0+a_1+a_2).$
Let $K=-(c_0(0)+c_1(0)+c_2(0)-a_0(0)+a_1(0)+a_2(0)).$
We see that the solution $\psi_n(y)$ 
of the equation $L\psi=0$ has the same behavior as
$e^\frac{Ky}{t}h(y,t),$ where $h$ is a holomorphic with respect to $t$
function such that $h(y,0)\ne0$ for the generic operator $L.$
It is easy to see that the Floquet multiplier $\mu$ in a neighborhood
of $Q$ has the same behavior as 
$\mu_3=e^{\int_0^T\frac{E(\xi)+F(\xi)+G(\xi)}{t}\,d\xi}.$ This implies that
$\mu$ can be presented 
as $e^\frac{M}{t}g(t),$ where $M$ is a
constant and $g(t)$ is a
holomorphic function such that $g(t)\ne0.$
$\Box$

Further we will consider only generic algebro-geometric operators.

Let us now study the divisor $\mathcal{D}(n,y)$ of the poles of
$\psi_n(y)$ on \mbox{$\Gamma\setminus\{Q,P^\pm_i\}.$} As it was proven
in Lemma~\ref{psi-function}, $\mathcal{D}(n,y)$ does not
depend on $y,$ thus we will write $\mathcal{D}(n).$ 

We need to consider the formal adjoint operator $L^+$ of $L.$
It is easy to see that
$$
(L^+\psi^+)_n=a_n\psi^+_n+(\psi^+_n)'+c_{n-1}\psi^+_{n-1}-(\psi^+_{n-1})'.
$$
Let us also consider the adjoint equation $L^+\psi^+=0.$ We can
as before consider Floquet solutions of this equation 
normalized by the condition $\psi^+_0(0)=1$
and see that $\psi^+_n(y)$ is a Baker-Akhiezer function on the corresponding
spectral curve.

\begin{Lemma}\label{adjoint}
The spectral curves of the equations $L\psi=0$
and $L^+\psi^+=0$ are isomorphic. We can 
identify
them and consider $\psi^+$ as a function on $\Gamma,$ in the sense that
a point $P=(\rho,\mu)$ corresponds to $\psi$  with Floquet multipliers 
$\rho$ and $\mu,$ and to $\psi^+$ with Floquet 
multipliers $\frac{1}{\rho}$ and
$\frac{1}{\mu}.$ In a neighborhood of 
the point $Q$ the function $\psi^+_n(y)$
can be presented as
$e^\frac{-Ky}{t}h^+(y,t),$
where $K$ is the same constant and 
$t$ is the same  local parameter at $Q$ as in Lemma~\ref{pointQ},
and $h^+(y,t)$ is a holomorphic with respect to $t$ 
function such that $h^+(y,0)\ne0.$
\end{Lemma}

\noindent{\bf Proof.} As was explained above, 
if $\psi$ is a Floquet solution
then the equations $L\psi=0$ can be rewritten as
$\Psi'(y)=A(y,\rho)\Psi(y),$ where 
$A(y,\rho)=-B^{-1}(\rho)C(y,\rho).$
In the same way if $\psi^+$ is a Floquet solution
with the Floquet multiplier $\frac{1}{\rho}$ then
the adjoint equation $L^+\psi^+=0$
can be written as $(\Psi^+(y))'=A^+(y,\rho)\Psi^+(y),$
where $A^+(y,\rho)=(B^{-1}(\rho))^T(C(y,\rho))^T.$

Let $\Phi(y,0)$ and $\Phi^+(y,0)$ be the resolvents of $A$ and $A^+.$
We know that
\begin{equation}\label{resolprop}
\frac{d}{dy}\Phi(y,0)=A(y)\Phi(y,0),
\quad
\frac{d}{dy}\Phi^+(y,0)=A^+(y)\Phi^+(y,0).
\end{equation}
Let $Q(y)=B^{-1}(\Phi^+(y,0))^TB\Phi(y,0).$ It is easy
to see that $Q(0)=I.$ It follows from~(\ref{resolprop}) that
$$
\frac{d}{dy}Q(y)=B^{-1}(\Phi^+(y,0))^TCB^{-1}B\Phi(y,0)-
$$
$$
-B^{-1}(\Phi^+(y,0))^TBB^{-1}C\Phi(y,0)=0.
$$
This implies that $Q(y)\equiv I.$ Hence $B^{-1}(\Phi^+(y,0))^TB$
is the inverse of $\Phi(T,0).$ This means that
the eigenvalues of $\Phi^+(T,0)$ and 
$\Phi(T,0)$ are inverse to each other. This implies that
the spectral curves of the equations $L\psi=0$
and $L^+\psi^+=0$ are isomorphic. 
We can  think that $\psi^+$ is a function on $\Gamma$ in the sense that
a point $P=(\rho,\mu)$ corresponds to $\psi$  with Floquet multipliers 
$\rho$ and $\mu,$ and to $\psi^+$ with Floquet 
multipliers $\frac{1}{\rho}$ and
$\frac{1}{\mu}.$ 

The statement about the behavior of $\psi^+$ in a neighborhood
of $Q$ can be proven in the same way as in 
Lemma~\ref{pointQ}.
$\Box$

\begin{Lemma} For any $k$ and $y$ the following identity holds
$$
\frac{d\rho}{\rho\sum_{n=k+1}^{k+N}(\psi_n(y)\psi^+_{n}(y)-%
\psi_n(y)\psi^+_{n-1}(y))}=
$$
$$
=\frac{d\mu}{\mu\int_{y}^{y+T}[c_k(y')\psi_{k+1}(y')\psi^+_k(y')+%
\psi'_{k+1}(y')\psi^+_k(y')]\,dy'}.
$$
\end{Lemma}

It is easy to see that this differential does not depend on $y$
since on the right hand side we integrate an expression periodic
in $y'.$ For an analogous reason this differential does not depend
on $k.$ We will denote this differential by $\Omega.$ 
Let us define
$$
R_\rho=\sum_{n=k+1}^{k+N}(\psi_n(y)\psi^+_{n}(y)-%
\psi_n(y)\psi^+_{n-1}(y)),
$$
and
$$
R_\mu=\int_{y}^{y+T}[c_k(y')\psi_{k+1}(y')\psi^+_k(y')+%
\psi'_{k+1}(y')\psi^+_k(y')]\,dy'.
$$
Thus $\Omega=\frac{d\rho}{\rho R_\rho}=\frac{d\mu}{\mu R_\mu}.$

\noindent{\bf Proof of the Lemma.} 
Let us multiply our equation
$$
a_n\psi_n(y,P)-\psi'_n(y,P)+c_n\psi_{n+1}(y,P)+\psi'_{n+1}(y,P)=0,
$$
by $\psi^+_n(y,\tilde{P}),$ and subtract the adjoint equation
$$
a_n\psi^+_n(y,\tilde{P})+(\psi^+_n(y,\tilde{P}))'+%
c_{n-1}\psi^+_{n-1}(y,\tilde{P})-(\psi^+_{n-1}(y,\tilde{P}))'=0,
$$
multiplied by $\psi_n(y,P).$ In the following formulas
we will presume that $\psi_n$ is taken at the point
$P,$ and $\psi^+_n$ is taken at the point $\tilde{P}.$  

Let us sum the resulting formula
$$
-(\psi_n\psi^+_n)'+c_n\psi_{n+1}\psi^+_n-c_{n-1}\psi_n\psi^+_{n-1}+%
\psi'_{n+1}\psi^+_n+\psi_n(\psi^+_{n-1})'=0
$$
from $k+1$ to $k+N.$
We can rewrite the resulting  formula using the fact that
$\psi$ and $\psi^+$ are Floquet solutions. We obtain
$$
-\sum_{n=k+1}^{k+N}(\psi_n\psi^+_n)'+%
\left(\frac{\rho(P)}{\rho(\tilde{P})}-1\right)c_k\psi_{k+1}\psi^+_k+%
$$
$$
+\sum_{n=k+1}^{k+N}(\psi_n\psi^+_{n-1})'+%
\left(\frac{\rho(P)}{\rho(\tilde{P})}-1\right)\psi'_{k+1}\psi^+_k=0.
$$
We integrate this formula and use again
the fact that
$\psi$ and $\psi^+$ are Floquet solutions. We obtain
$$
-\left(\frac{\mu(P)}{\mu(\tilde{P})}-1\right)%
\sum_{n=k+1}^{k+N}\psi_n\psi^+_n+%
\left(\frac{\rho(P)}{\rho(\tilde{P})}-1\right)%
\int_y^{y+T}c_k\psi_{k+1}\psi^+_k\,dy'+
$$
$$
+\left(\frac{\mu(P)}{\mu(\tilde{P})}-1\right)%
\sum_{n=k+1}^{k+N}\psi_n\psi^+_{n-1}+%
\left(\frac{\rho(P)}{\rho(\tilde{P})}-1\right)%
\int_y^{y+T}\psi'_{k+1}\psi^+_k\,dy'=0.
$$
Taking a limit when $\tilde{P}$ tends to $P$ we obtain
the formula in the statement.
$\Box$

\begin{Lemma}\label{commonzeroes}
The functions $R_\rho$ and $R_\mu$
have no common zeroes in \mbox{$\Gamma\setminus\{Q,P^\pm_i\}.$}
\end{Lemma}

\noindent{\bf Proof} is similar to the proof of the previous Lemma.
Let us consider the curves $a_n(y,t),$ $c_n(y,t),$ $\rho(t),$ $\mu(t),$
$\psi_n(y,t)$ such that $\psi_n(y,t)$ is a solution of the equation 
$L(t)\psi=0$
with Floquet multipliers $\rho(t)$ and $\mu(t)$ and
$a_n(y,0)=a_n(y)$ etc.

Let us multiply
$$
a_n(y,t)\psi_n(y,t)-\psi'_n(y,t)+c_n(y,t)\psi_{n+1}(y,t)+\psi'_{n+1}(y,t)=0,
$$
by $\psi^+_n(y),$ and subtract the adjoint equation
$$
a_n(y)\psi^+_n(y)+(\psi^+_n(y))'+c_{n-1}(y)\psi^+_{n-1}(y)-%
(\psi^+_{n-1}(y))'=0,
$$
multiplied by $\psi_n(y,t).$ Let us sum and integrate as before.
Taking the derivative with respect to $t$ at $t=0$ we obtain
$$
\sum_{n=k+1}^{k+N}\int_y^{y+T}\left[%
\frac{\partial a_n}{\partial t}(y',0)\psi_n(y')\psi^+_n(y')+%
\frac{\partial c_n}{\partial t}(y',0)\psi_{n+1}(y')\psi^+_n(y')%
\right]\,dy'-
$$
$$
-\frac{\frac{\partial\mu}{\partial t}(0)}{\mu}%
\sum_{n=k+1}^{k+N}(\psi_n(y)\psi^+_n(y)-\psi_n(y)\psi^+_{n-1}(y))+
$$
$$
+\frac{\frac{\partial\rho}{\partial t}(0)}{\rho}%
\int_y^{y+T}[c_k(y')\psi_{k+1}(y')\psi^+_k(y')+%
\psi'_{k+1}(y')\psi^+_k(y')]\,dy'=0.
$$
If $R_\rho$ and $R_\mu$ have a common zero then in this point
$$
\sum_{n=k+1}^{k+N}\int_y^{y+T}%
\left[\frac{\partial a_n}{\partial t}(y',0)\psi_n(y')+%
\frac{\partial c_n}{\partial t}(y',0)%
\psi_{n+1}(y')\right]\psi^+_n(y')\,dy'=0
$$
for any $\frac{\partial a_n}{\partial t}(y,0)$ 
$\frac{\partial c_n}{\partial t}(y,0).$ 
It follows that for any $n$ and $y$ we have $\psi^+_n(y)=0,$
but this contradicts the normalization condition $\psi^+_0(0)=1.$
$\Box$

It follows from Lemma~\ref{pointQ} that $\frac{d\mu}{\mu}$ 
has a pole of order 2
at $Q.$ Let us consider $\Omega=\frac{d\mu}{\mu R_\mu},$ 
where $R_\mu$ is
written for example for $k=0$ and $y=0:$
$$
R_\mu=\int_{0}^{T}[c_0(y')\psi_{1}(y')+%
\psi'_{1}(y')]\psi^+_0(y')\,dy'.
$$
It follows from the behavior of $\psi$ and $\psi^+$ at the point $Q$
described in Lemmas~\ref{pointQ} and~\ref{adjoint} that $\Omega$
has a pole of first order at $Q.$

From Lemmas~\ref{Pplus} and~\ref{Pminus} we know the structure
of poles and zeroes of $\psi_n$ at the points $P^\pm_i.$ 
We can prove in the same way that $\psi^+_0(y)$ has no zeroes or poles
at $P^\pm_i.$ It follows that in these points $\Omega$ has only
a first order pole at $P^+_1.$

It is easy to see that $\frac{d\rho}{\rho}$ and $\frac{d\mu}{\mu}$
have no poles in $\Gamma\setminus\{Q,P^\pm_i\}.$ 
It follows from Lemma~\ref{commonzeroes} that
$\Omega$ is holomorphic in $\Gamma\setminus\{Q,P^\pm_i\}.$ 

For a generic operator 
the functions $a_n\psi_n-\psi'_n$ and $c_{n-1}\psi_n+\psi'_n$ have
in $\Gamma\setminus\{Q,P^\pm_i\}$ the same poles as $\psi_n.$ It follows
from the equation $L\psi=0$
that $\mathcal{D}(n)$ does not depend on $n$ and we will write 
$\mathcal{D}.$
We can give the same argument with the divisor $\mathcal{D}^+$ corresponding
to $\psi^+.$ Since for a generic operator $d\rho$ and $d\mu$ have no common 
zeroes, we obtain the following Lemma.

\begin{Lemma} $(\Omega)=-Q-P^+_1+\mathcal{D}+\mathcal{D}^+.$
\end{Lemma}

Let $g$ be the genus of $\Gamma.$ 
Since the canonical class has degree $2g-2,$
we see that $\mathcal{D}$ is an effective divisor of 
degree $g.$

We can now summarize the results of this Section in the 
following Theorem.

\begin{Theorem}\label{direct}
 Let $L$ be a generic algebro-geometric operator. Then its
spectral curve $\Gamma$ is a compact Riemann 
surface. Let $g$ be the genus of $\Gamma.$
The Floquet solution $\psi_n(y)$ is a Baker-Akhiezer function on $\Gamma.$
There are points $P^\pm_i,$ $i=1,\dots,N,$ such that $\psi_n(y)$
has a zero of order $\left[\frac{n-i}{N}\right]+1$ at  $P^+_i$
and a pole of the same order at $P^-_i.$ There is a point $Q$ such that
the function $\psi_n(y)$ is meromorphic in $\Gamma\setminus\{Q\}.$
The function $\psi_n(y)$ can be presented 
in a neighborhood of $Q$ as $e^\frac{Ky}{t}h_n(y,t),$
where $K$ is a constant, $t$ is a local parameter at $Q$
and $h_n(y,t)$ is a function that is holomorphic with respect to $t,$ 
such that $h(y,0)\ne0.$ The divisor
$\mathcal{D}$ of poles of $\psi_n(y)$ in $\Gamma\setminus\{Q,P^\pm_i\}$
is an effective divisor of degree $g$ and
does not depend on $n$ or $y.$
\end{Theorem}

\subsubsection{Inverse spectral problem}

Let us consider a non-singular curve $\Gamma$ of genus $g,$ labelled points
$P^\pm_i,$ $i=1,\dots,N,$ and $Q$ on the curve $\Gamma,$
and also a divisor
$\mathcal{D}=R_1+\dots+R_g.$ Let us also fix a $1$-jet $[\lambda]_1$
of a local parameter at the point $Q,$ i.e. a local parameter at the point
$Q$ up to a transformation $\tilde{\lambda}=\lambda+O(\lambda^2).$ 
The set $(\Gamma,P^\pm_i,Q,[\lambda]_1,\mathcal{D}=R_1+\dots+R_g)$ is 
called the {\em spectral data.}

As we have seen in Theorem~\ref{direct}, we can build spectral data
starting from a generic periodic algebro-geometric operator $L$ 
(we should take $\lambda=\frac{t}{K}$ ).
Our goal is to
prove the inverse theorem. 

\begin{Theorem} Let $(\Gamma,P^\pm_i,Q,[\lambda]_1,\mathcal{D}=R_1+\dots+R_g)$ 
be spectral data with a generic 
$\mathcal{D}.$
There exists a Baker-Akhiezer function $\psi_n(y)$ defined on 
$\Gamma$ and depending on two parameters $n\in\mathbb{N}$ and $y\in\mathbb{R}$ 
such that
\begin{enumerate}
\item $\psi_0(0)=1.$
\item The function $\psi_n(y)$
has a zero of order 
$\left[\frac{n-i}{N}\right]+1$
at $P^+_i$ and a pole of the same order 
at $P^-_i.$
\item The function $\psi_n(y)$ is meromorphic in $\Gamma\setminus\{Q\}.$
\item The poles of the function $\psi_n(y)$ in $\Gamma\setminus\{P^\pm_i,Q\}$
can be only first-order poles at points $R_i.$
\item The product $\psi_n(y)e^{-\frac{y}{\lambda}}$
is a holomorphic function in a neighborhood of $Q$ and it is not equal to zero
at $Q.$
\item There exists an operator $L$ of the form~(\ref{semi-discrete})
such that $L\psi=0.$
\item The function $\psi_n(y)$ and $L$ are defined by spectral data uniquely
up to a gauge transformation.
\end{enumerate}
\end{Theorem}

\noindent{\bf Proof.} We will use the same notation and conventions as 
in the paper~\cite{D}. Let us choose a basis in cycles.
Let $\Omega_{P^+_iP^-_i}$ be the Abelian differential
of the third kind with poles at points
$P^+_i$ and $P^-_i$ and residues $+1$ and $-1$ respectively.
Let $\Omega$ be the Abelian differential of the second kind
with a pole of second order at $Q$ and the Laurent series expansion 
$\Omega=(-\frac{1}{\lambda^2}+\dots)\,d\lambda.$
We recall that $a$-periods of $\Omega$ and $\Omega_{P^+_iP^-_i}$ 
are equal to zero.
It is easy to see that $\Omega$ depends only
on the $1$-jet $[\lambda]_1.$
Let $U$ and $U_{P^+_iP^-_i}$ denote the vectors
of $b$-periods of $\Omega$ and $\Omega_{P^+P^-}$ respectively. 

Let us define $P^\pm_i$ for all $i$ by periodicity: $P^\pm_{i+N}=P^\pm_i.$
Let us introduce the sign $\sumprime$ in the following way:
$$
\sumprime_{i=1}^n a_i=\left\{%
\begin{array}{lcl}
\sum\limits_{i=1}^{n}a_i&\mbox{if}& n>1,\\
0&\mbox{if}&n=0,\\
-\sum\limits_{i=n}^{-1}a_i&\mbox{if}&n<0.
\end{array}
\right.
$$

It is easy to check that the function
\begin{equation}\label{psitheta}
\exp(\int_{P_0}^Py\Omega+%
\sumprime\limits_{i=1}^{n}\Omega_{P^+_iP^-_i})%
\frac{\Theta(A(P)+yU+\sumprime\limits_{i=1}^{n}U_{P^+_iP^-_i}%
-A(\mathcal{D})-\mathcal{K})}%
{\Theta(A(P)-A(\mathcal{D})-\mathcal{K})},
\end{equation}
depending on two parameters $n$ and $y,$
satisfies the conditions (1)--(5) from the statement of this theorem.
Hence the existence of $\psi_n(y)$ with properties (1)--(5) is proven.

Let us now prove that any function $\psi_n(y)$ 
satisfying the conditions (1)--(5) has the form
\begin{equation}\label{psirtheta}
\psi_n(y,P)=r_n(y)\exp(\int_{P_0}^Py\Omega+%
\sumprime\limits_{i=1}^{n}\Omega_{P^+_iP^-_i})\times%
\end{equation}
$$
\times\frac{\Theta(A(P)+yU+\sumprime\limits_{i=1}^{n}U_{P^+_iP^-_i}%
-A(\mathcal{D})-\mathcal{K})}%
{\Theta(A(P)-A(\mathcal{D})-\mathcal{K})},
$$
where $r_n(y)$ are constants (i.e. do not depend on a point on $\Gamma$)
such that $r_0(0)=1.$

Indeed, the ratio of a function $\psi_n(y)$ and the function~(\ref{psitheta})
is a meromorphic function on $\Gamma$ with a pole divisor
$\mathcal{D}-yU-nU_{P^+P^-},$ which is generic. 
It follows that the space of such meromorphic functions has
dimension $1,$ i.e. this ratio is a constant function
on $\Gamma,$ depending on the parameters $n$ and $y.$ 
Let us denote this constant as $r_n(y).$ It follows from the condition (1)
that $r_0(0)=1.$

Let us now construct the operator $L.$ We already have the function $\psi_n(y)$
given by the formula~(\ref{psirtheta}). 
Let us consider the equation $L\psi=0$
as an equation for unknown $a_n(y),\dots,d_n(y):$
\begin{equation}\label{abcdeq}
a_n(y)\psi_n+b_n(y)\psi'_n+c_n(y)\psi_{n+1}+d_n(y)\psi'_{n+1}=0.
\end{equation}
Let us take the Laurent series expansion of~(\ref{abcdeq}) at the point
$P^-_{n+1}.$ The first term of this expansion gives us the identity
$$
c_n(y)r_{n+1}(y)\exp\int_{P_0}^{P^-_{n+1}}\!\!\!\!\!\!y\Omega%
\frac{\Theta(A(P^-_{n+1})+yU+\sumprime\limits_{i=1}^{n+1}U_{P^+_iP^-_i}-%
A(\mathcal{D})-\mathcal{K})}%
{\Theta(A(P^-_{n+1})-A(\mathcal{D})-\mathcal{K})}+
$$
$$
+d_n(y)\Biggl[r'_{n+1}(y)\exp\int_{P_0}^{P^-_{n+1}}\!\!\!\!\!\!y\Omega%
\frac{\Theta(A(P^-_{n+1})+yU+\sumprime\limits_{i=1}^{n+1}U_{P^+_iP^-_i}-%
A(\mathcal{D})-\mathcal{K})}%
{\Theta(A(P^-_{n+1})-A(\mathcal{D})-\mathcal{K})}+
$$
$$
+r_{n+1}(y)\int_{P_0}^{P^-_{n+1}}\!\!\!\!\!\Omega%
\exp\int_{P_0}^{P^-_{n+1}}\!\!\!\!\!\!y\Omega%
\frac{\Theta(A(P^-_{n+1})+yU+\sumprime\limits_{i=1}^{n+1}U_{P^+_iP^-_i}-%
A(\mathcal{D})-\mathcal{K})}%
{\Theta(A(P^-_{n+1})-A(\mathcal{D})-\mathcal{K})}+
$$
$$
+r_{n+1}(y)\exp\int_{P_0}^{P^-_{n+1}}\!\!\!\!\!\!y\Omega%
\frac{\frac{\partial}{\partial y}%
\Theta(A(P^-_{n+1})+yU+\sumprime\limits_{i=1}^{n+1}U_{P^+_iP^-_i}-%
A(\mathcal{D})-\mathcal{K})}%
{\Theta(A(P^-_{n+1})-A(\mathcal{D})-\mathcal{K})}\Biggr]=0.
$$
This implies that
\begin{equation}\label{cd}
\frac{c_n(y)}{d_n(y)}=%
-\frac{r'_{n+1}(y)}{r_{n+1}(y)}-%
\int_{P_0}^{P^-_{n+1}}\!\!\!\!\!\Omega-
\end{equation}
$$
-\frac{\partial}{\partial y}\ln%
\Theta(A(P^-_{n+1})+yU+\sumprime\limits_{i=1}^{n+1}U_{P^+_iP^-_i}-%
A(\mathcal{D})-\mathcal{K}).
$$
In the same way the term of order $\lambda^{-1}$ in 
the Laurent series expansion
of the equation $(L\psi)_ne^{-\frac{y}{\lambda}}=0$ at 
$Q$ gives us the formula
\begin{equation}\label{db}
\frac{d_n(y)}{b_n(y)}=-\frac{r_n(y)}{r_{n+1}(y)}%
\exp(-\int_{P_0}^Q\Omega_{P^+_{n+1}P^-_{n+1}})\times
\end{equation}
$$
\times\frac{\Theta(A(Q)+yU+\sumprime\limits_{i=1}^nU_{P^+_iP^-_i}-%
A(\mathcal{D})-\mathcal{K})}%
{\Theta(A(Q)+yU+\sumprime\limits_{i=1}^{n+1}U_{P^+_iP^-_i}-%
A(\mathcal{D})-\mathcal{K})}.
$$
Considering the Laurent series expansion of~(\ref{abcdeq}) 
at the point $P^+_{n+1}$
we obtain the formula
\begin{equation}\label{ab}
\frac{a_n(y)}{b_n(y)}=%
-\frac{r'_{n}(y)}{r_{n}(y)}-%
\int_{P_0}^{P^+_{n+1}}\!\!\!\!\!\Omega-
\end{equation}
$$
-\frac{\partial}{\partial y}
\Theta(A(P^+_{n+1})+yU+\sumprime\limits_{i=1}^{n}U_{P^+_iP^-_i}-%
A(\mathcal{D})-\mathcal{K}).
$$
If $a_n(y),\dots,d_n(y)$ satisfy 
identities~(\ref{cd}), (\ref{db}) and~(\ref{ab}) then
the equation~(\ref{abcdeq}) holds at any point on $\Gamma.$ 
To prove this, let us 
consider the function 
$$
\varphi_n(y)=-\frac{b_n(y)}{a_n(y)}\psi'_n(y)-%
\frac{c_n(y)}{a_n(y)}\psi_{n+1}(y)-%
\frac{d_n(y)}{a_n(y)}\psi'_{n+1}(y).
$$
It follows from the formulas~(\ref{cd}) and~(\ref{db})
that $\varphi_n(y)$ satisfies the same conditions (1)--(5) as $\psi_n(y).$
As explained before, it follows that these two functions are proportional.
The formula~(\ref{ab}) means that the coefficient of
proportionality is equal to $1,$ i.e. 
$\varphi_n(y)\equiv\psi_n(y).$ This implies
that the equation~(\ref{abcdeq}) holds at any point on $\Gamma.$ 

The identities~(\ref{cd}), (\ref{db}) and~(\ref{ab})
determine $a_n(y),\dots,d_n(y)$
up to a multiplication by
a constant depending on $n$ and $y.$ It is easy to check that 
this fact together with
the fact that $\psi_n(y)$ are determined up to constants $r_n(y)$ means exactly
that $L$ and $\psi$ are defined up to a gauge transformation. $\Box$

We should remark that operators obtained by this Theorem from 
spectral data are not
necessarily gauge equivalent to periodic ones.

\subsection{Algebro-geometric spectral theory of two-dimensional 
discrete hyperbolic Schr\"odinger operators}

Let us consider a discrete operator $L$ of the form~(\ref{discrete})
and the corresponding equation $L\psi=0$~(\ref{discreteequation})
with the following matrix of periods
\begin{equation}\label{T}
{\mathfrak T}=\left(\begin{array}{cc} P&R\\S&T\end{array}\right),
\end{equation}
i.e.
\begin{gather*}
a_{n+P,m+R}=a_{n+
S,m+T}=a_{n,m}, b_{n+P,m+R}=b_{n+S,m+T}=b_{n,m},\\
c_{n+P,m+R}=c_{n+S,m+T}=c_{n,m}, d_{n+P,m+R}=d_{n+S,m+T}=d_{n,m},
\end{gather*}

Let $\Delta=\det(\mathfrak T)>0$. The matrix
$\mathfrak{T}$ defines a sub-lattice of $\mathbb Z\times
\mathbb Z$. We call it a period sub-lattice.
Let us fix two specific choices of basis of the
period sub-lattice.

\begin{Lemma}
For the period sub-lattice
given by the matrix ${\mathfrak T}$ there exists
a unique basis such that  the corresponding matrix
$$
\mathfrak T_1=
\left(\begin{array}{cc}\tilde\delta&0\\-\zeta&\delta\end{array}\right)
$$
satisfies conditions $\delta=(R,T)$, $\tilde\delta=\Delta/\delta$,
$0\leqslant\zeta<\tilde\delta.$

There also exists a unique basis such that
the corresponding matrix
$$
\mathfrak T_2=
\left(\begin{array}{cc}\varepsilon&-\xi\\0&\tilde\varepsilon\end{array}\right)
$$
satisfies the conditions $\varepsilon=(P,S)$,
$\tilde\varepsilon=\Delta/\varepsilon$, $\Delta=\det(\mathfrak T)$,
$0\leqslant\xi<\tilde\varepsilon$. 
\end{Lemma}

\noindent{\bf Proof} is a direct calculation. $\Box$

We consider only the case when 
$\delta<\Delta$, $\varepsilon<\Delta$. It is a reasonable
assumption. Indeed, if, for example, $\delta=\Delta$ then
$\tilde{\delta}=1.$ It means that one of the periods is equal to $1.$
This degenerate case is of little interest.

The main object of our interest is the Floquet solution
$\psi$ of the equation $L\psi=0.$
Since we have two choices of basis of the period sub-lattice,  
we can define two pairs of corresponding Floquet multipliers of the 
function $\psi_{n,m}$:
\begin{gather}
\psi_{n+\tilde\delta,m}=\nu_1\psi_{n,m},\quad
\psi_{n-\zeta,m+\delta}=\mu_1\psi_{n,m},\label{per1}\\
\psi_{n,m+\tilde\varepsilon}=\mu_2\psi_{n,m},\quad
\psi_{n+\varepsilon,m-\xi}=\nu_2\psi_{n,m}\label{per2}.
\end{gather}

Since $\mathfrak T_1\mathfrak T_2^{-1}\in SL_2(\mathbb Z)$, it follows that 
$1-\tilde\xi\tilde\zeta=\varkappa\tilde\Delta,$
where $\xi=\tilde\xi\varepsilon$, $\zeta=\tilde\zeta\delta$,
$\Delta=\tilde\Delta\varepsilon\delta$, $\varkappa\in\mathbb Z_+$.
Thus we can express  $\nu_1,\mu_1$ in terms of $\nu_2,\mu_2$ 
and vice versa by the formulas
\begin{align}
\nu_1&=\nu_2^{\tilde\delta}\mu_2^{\tilde\zeta}, &
\mu_1&=\nu_2^{-\tilde\zeta}\mu_2^{\varkappa}, \label{1-2} \\
\nu_2&=\nu_1^\varkappa\mu_1^{-\tilde\zeta}, &
\mu_2&=\nu_1^{\tilde\xi}\mu_1^{\tilde\delta}. \label{2-1}
\end{align}

\subsubsection{The inverse spectral problem}
The inverse spectral problem for the discrete 
operators was solved by Krichever~\cite{K}.
Let us recall (in our notation) 
the main theorem of the paper~\cite{K}

\begin{Theorem}\label{mainq1} Let the matrix $\mathfrak T$~(\ref{T})
define a period sub-lattice  (and hence $\mathfrak{T}_1,$ $\mathfrak{T}_2$).
Let $C$ be a curve of genus $g=\Delta-\delta-\epsilon+1$ and 
$P^\pm_n, Q^\pm_n$ be points on $C$ such that 
\begin{equation}\label{periodPQ}
P^\pm_n=P^\pm_{n'},\,\, Q^\pm_k=Q^\pm_{k'}\quad%
\mbox{for}\quad n-n'=0\!\!\!\mod\delta,\,k-k'=0\!\!\!\mod\varepsilon.
\end{equation}
Let $\mathcal{D}$ be a generic effective divisor of degree $g.$

Let one of the following  equivalent conditions be satisfied
\begin{align}\label{divper1}
\sum_{i=1}^{\tilde\delta}(Q^+_i-Q^-_i)&\sim(\nu_1),&
\sum_{j=1}^\delta(P^+_i-P^-_i)-\sum_{i=1}^{\zeta}(Q^+_i-Q^-_i)&\sim
(\mu_1),\\ \sum_{i=1}^{\tilde\epsilon}(P^+_i-P^-_i)&\sim (\mu_2),&
\label{divper2}
\sum_{j=1}^\epsilon(Q^+_i-Q^-_i)-\sum_{i=1}^{\xi}(P^+_i-P^-_i)&
\sim(\nu_2),
\end{align}
where $\nu_i,\mu_i$ are meromorphic functions on the curve.

Then the following statements hold.
\begin{enumerate}
\item The space
 $$
\mathcal L_{m,n}(\mathcal D)=\mathcal
L(\sumprime_{i=1}^{m}(Q^-_i-Q^+_i)+
\sumprime_{i=1}^{n}(P^-_i-P^+_i)+\mathcal{D}).
$$
is one-dimensional.
\item For any choice of nonzero functions 
$\psi_{m,n}\in\mathcal L_{m,n}(\mathcal D)$
satisfying the Floquet conditions (\ref{per1}),
(\ref{per2}) there exists  an operator $L$  with period matrix 
$\mathfrak T$ such that $L\psi_{n,m}=0.$
The operator $L$ and the function $\psi_{n,m}$ are unique up to 
a gauge transformation~(\ref{gaugediscrete}).
\end{enumerate}
\end{Theorem}

The explicit formulas for $L$ and $\psi_{n,m}$ 
in terms of theta-functions can be found in~\cite{K}.

This means that given the spectral data 
$(\Gamma,P^\pm_n,Q^\pm_n,\mathcal{D})$ satisfying the 
conditions~(\ref{periodPQ}), (\ref{divper1}) and~(\ref{divper2}),
one can reconstruct $L$ and $\psi_{nm}$
uniquely up to a gauge transformation in terms of theta-functions.

Counting of parameters performed in the paper~\cite{K}  shows that
varying the curve $C$ and the divisor $\mathcal{D}$ we can get a dense
subset of the set of periodic discrete operators. In this paper this
fact is proved effectively by solving direct spectral problem. 

\subsubsection{The direct spectral problem}

For discrete operators the direct spectral problem 
can be solved explicitly. Let us consider a generic discrete operator
$L$ with a period matrix $\mathfrak{T}$~(\ref{T})
and the equation $L\psi=0.$ As in the semi-discrete case, we
consider the Floquet solution. 

\begin{Theorem}\cite{K,O} The possible pairs of Floquet
multipliers form a curve which can be compactified. 
The compactified curve $\Gamma$ (called a spectral curve)
has genus $g=\Delta-\delta-\varepsilon+1.$  There exist an effective
divisor $\mathcal{D}$ of degree $g$ and points  $P_k^\pm$, $Q^\pm_l$
satisfying the conditions~(\ref{periodPQ}), (\ref{divper1}) 
and~(\ref{divper2}), such
that the function $\psi_{n,m}$ normalized by the condition
$\psi_{0,0}=1$ is meromorphic function belonging to the space
$\mathcal{L}_{n,m}(\mathcal{D}).$
\end{Theorem}

\noindent{\bf Proof.}
Let us introduce some notations:
\begin{align*}
A_{.,j}&=\prod_{i=1}^{\tilde\delta}a_{i,j}&
B_{.,j}&=\prod_{i=1}^{\tilde\delta}b_{i,j}&
C_{.,j}&=\prod_{i=1}^{\tilde\delta}c_{i,j}&
D_{.,j}&=\prod_{i=1}^{\tilde\delta}d_{i,j}\\
A_{i,.}&=\prod_{j=1}^{\tilde\varepsilon}a_{i,j}&
B_{i,.}&=\prod_{j=1}^{\tilde\varepsilon}b_{i,j}&
C_{i,.}&=\prod_{j=1}^{\tilde\varepsilon}c_{i,j}&
D_{i,.}&=\prod_{j=1}^{\tilde\varepsilon}d_{i,j}. 
\end{align*}

Let the matrix $M$ be the matrix of the linear system $L\psi=0$ written
in some basis. In more details, let us put the coefficients of the equation
$(L\psi)_{i,j}=0$ in the $i+j\tilde\delta+1$-th  row
($0\leqslant i<\tilde\delta$, $0\leqslant j<\delta$) of $M,$
where $\psi_{k,l}$ are written in the form
 $\psi_{k,l}=\nu_1^\alpha\mu_1^\beta\psi_{k',l'}$, $0\leqslant
k'<\tilde\delta$, $0\leqslant l'<\delta.$
The $1+i+j\tilde\delta$-th column 
($0\leqslant i<\tilde\delta$, $0\leqslant j<\delta$)
of $M$  corresponds to $\psi_{i,j}$. 
The constructed matrix $M$ has a block triangular
structure and the equation $R(\mu_1,\nu_1):=\det M=0$ is the equation
of an affine part of the spectral curve.

We also need a matrix $\hat M$ constructed in an analogous way. 
Let us put the coefficients of the
equation $(L\psi)_{i,j}=0$ in the 
$i\tilde\varepsilon+j+1$-th row ($0\leqslant i<\varepsilon$, $0\leqslant
j<\tilde\varepsilon$) of the matrix $\hat M,$ 
where $\psi_{k,l}$ are presented in
the form  $\psi_{k,l}=\nu_2^\alpha\mu_2^\beta\psi_{k',l'},$
$0\leqslant k'<\varepsilon$, $0\leqslant l'<\tilde\varepsilon.$  The
$1+i\tilde\varepsilon+j$-th column ($0\leqslant i<\varepsilon$, $0\leqslant
j<\tilde\varepsilon$) corresponds to $\psi_{i,j}$. Let
$\hat{R}(\nu_2,\mu_2)=\det\hat{M}$.

Reformulating Lemma~1 from the paper~\cite{O} we get the following Lemma.

\begin{Lemma}\label{osntec} 
For any operator $L$ the following identities hold.
$$
R(\mu_1,\nu_1)=\hat R(\mu_2,\nu_2),
$$
$$
R(\nu_1,\mu_1)= \sum_{\genfrac{}{}{0pt}{}
{\genfrac{}{}{0pt}{}{0\leqslant 
i\tilde\Delta-\tilde\zeta j\leqslant\tilde\varepsilon}
{0\leqslant
\tilde\xi i+\varkappa j\leqslant\varepsilon+\xi}} {\genfrac{}{}{0pt}{}%
{0\leqslant i\leqslant%
\delta+\zeta}{0\leqslant j\leqslant\tilde\delta}}} r_{i,j}\nu_1^i\mu_1^j,\quad
\hat R(\nu_2,\mu_2)= \sum_{\genfrac{}{}{0pt}{}
{\genfrac{}{}{0pt}{}{0\leqslant i\varkappa+\tilde\zeta
j\leqslant\delta+\varepsilon} {0\leqslant -\tilde\xi i+\tilde\delta
j\leqslant\tilde\delta+\xi}} {\genfrac{}{}{0pt}{}{0\leqslant
i\leqslant\tilde\varepsilon}{0\leqslant j\leqslant\varepsilon+\zeta}}} \hat
r_{i,j}\nu_2^i\mu_2^j,
$$
$$
\sum_{i=0}^{\delta+\zeta}
r_{i,0}\nu_1^i=\prod_{j=0}^{\delta-1}\left(B_{.,j}\nu_1-
\left(-1\right)^{\tilde\delta}A_{.,j}\right),
$$
$$
\sum_{i=0}^{\delta+\zeta}             
r_{i,\tilde\delta}\nu_1^i=\nu_1^\zeta
\prod_{j=0}^{\delta-1}\left(D_{.,j}\nu_1-
\left(-1\right)^{\tilde\delta}C_{.,j}\right),
$$
$$
\sum_{j=0}^{\epsilon+\xi} \hat
r_{0,j}\mu_2^j=\prod_{i=0}^{\epsilon-1}\left(C_{i,.}\mu_2-
\left(-1\right)^{\tilde\epsilon}A_{i,.}\right),
$$
$$
\sum_{j=0}^{\epsilon+\xi} \hat r_{\tilde\epsilon,j}\mu_2^j=
\mu_2^\xi\prod_{i=0}^{\epsilon-1}\left(D_{i,.}\mu_2-
\left(-1\right)^{\tilde\epsilon}B_{i,.}\right).
$$
\end{Lemma}

\noindent{\bf Proof} is by a direct calculation. $\Box$

From this Lemma we see that for a generic operator $L$ the
natural compactification of the affine part of the spectral curve
$\{(\nu_1,\mu_1)| R(\nu_1,\mu_1)=0, \nu_1\mu_1\ne 0\}$ is a curve
of genus $\Delta-\delta-\epsilon+1$. Compactifying this curve
we add four groups of points $P_i^{\pm}$, $Q_i^{\pm}$. In the
coordinates  $\nu_1,\mu_1$ we have
$$
P_i^+=((-1)^{\tilde\delta}\frac{A_{.,j}}{B_{.,j}},0),\quad
P_i^-=((-1)^{\tilde\delta}\frac{C_{.,j}}{D_{.,j}},\infty);
$$
and in the coordinates $\nu_2,\mu_2$ we have
$$
Q_i^+=(0,(-1)^{\tilde\epsilon}\frac{A_{i,.}}{C_{i,.}}),\quad
Q_i^-=(\infty,(-1)^{\tilde\epsilon}\frac{B_{i,.}}{D_{i,.}}).
$$

It is easy to see that
$$
\frac{\psi_{n+1,m}}{\psi_{n,m}}= \frac{\frac{\partial
R}{\partial b_{n,m}}}{\frac{\partial R}{\partial a_{n,m}}},\quad
\frac{\psi_{n,m+1}}{\psi_{n,m}}=\frac{\frac{\partial R}{\partial c_{n,m}}}%
{\frac{\partial R}{\partial a_{n,m}}}.
$$ 
Using this formula together
with Lemma~\ref{osntec} we can prove that 
there exists an effective divisor $\mathcal{D}$ of degree $g$ such that
$\psi\in\mathcal{L}_{n,m}(\mathcal{D}).$ For the proof
see Lemma~2 of the paper~\cite{O}.

To get more information about the divisor $\mathcal{D}$ we need
the formal adjoint operator
$$
(L^+\psi^+)_{n,m}=a_{n,m}\psi^+_{n,m}+b_{n-1,m}\psi^+_{n,m}+
c_{n,m-1}\psi^+_{n,m-1}+d_{n-1,m-1}\psi^+_{n-1,m-1}.
$$
We are interested in a Floquet solution of the equation $L^+\psi^+=0$
such that
\begin{align}
\psi^+_{n+\tilde\delta,m}&=(\nu_1)^{-1}\psi^+_{n,m}&
\psi^+_{n-\zeta,m+\delta}&=(\mu_1)^{-1}\psi^+_{n,m} \label{per1'}\\
\psi^+_{n,m+\tilde\varepsilon}&=(\mu_2)^{-1}\psi^+_{n,m}&
\psi^+_{n+\varepsilon,m-\xi}&=(\nu_2)^{-1}\psi^+_{n,m}
\end{align}

It is easy to see that the $1+i+j\tilde\delta$-th row of the matrix
$M^T$ corresponds to the equation $(L^+\psi^+)_{i,j}=0$ 
($0\leqslant i<\tilde\delta$, $0\leqslant j<\delta$), 
where $\psi^+$ is written in the
form  $\psi^+_{m,n}=\nu_1^\alpha\mu_2^\beta\psi^+_{n',m'}$, with
$0\leqslant n'<\tilde\delta$, $0\leqslant m'<\delta.$ 
The $1+i+j\tilde\delta$-th column of $M^T$ corresponds to $\psi^+_{i,j}$. 
The operator $L^+$ is of a form similar to the form of $L.$ 
Moreover, the spectral curves
of these operators are isomorphic.
The function $\psi^+$ belong to the space 
$\mathcal{L}_{n,m}(\mathcal{D}^+)$ for 
some effective divisor $\mathcal{D}^+.$ 
Using this observation we get (see Lemma~5 of the paper~\cite{O})
the following

\begin{Lemma}\label{dif}  Let $L$ be a generic operator.  
Then the differential
\begin{gather*}
\Omega=\frac{\tilde{\Delta}d\nu_1}{\mu_1\nu_1 R_{\mu_1}},
\end{gather*}
where $\tilde{\Delta}$ is the determinant of the algebraic
complement of the element $M_{11}$ of the matrix $M$, satisfies
the identity
\begin{gather*}
(\Omega)=-P^+_0-Q^+_0+\mathcal D+\mathcal D^+.
\end{gather*}
It follows that $|\mathcal D|=|\mathcal{D}^+|=\Delta-\delta-\epsilon+1=g$.
\end{Lemma}

This lemma completes the proof of the Theorem. $\Box$

\section{Spectral properties of the Laplace transformations}

\subsection{Spectral properties of the 
Laplace transformations of algebro-geometric 
two-dimensional semi-discrete hyperbolic Schr\"odinger operators}

Since the Laplace transformations act on 
gauge equivalence classes of operators, and
operators can be obtained from spectral data, we can ask how to describe the 
Laplace transformations in terms of spectral data. 
It turns out that the Laplace 
transformations are shifts on the  Jacobian of a spectral curve.

\begin{Theorem}\label{laplacesp}
The Laplace transformations 
act on spectral data in the following
way: $\Gamma,$ $P^+_i,$ $Q,$ and $[\lambda]_1$ 
are not changing. 
The points
$P^-_i$ and the divisor $\mathcal{D}$ are changing according to the rule
$$
\tilde{\mathcal{D}}=\mathcal{D}+P^-_1-Q,\quad \tilde{P}^-_i=P^-_{i+1}
$$
for the Laplace transformation of the first type and according to
the rule
$$
\tilde{\mathcal{D}}=\mathcal{D}-P^-_N+Q,\quad \tilde{P}^-_i=P^-_{i-1}
$$
for the Laplace transformation of the second type. 
\end{Theorem}

When we write a formula like 
$\tilde{\mathcal{D}}=\mathcal{D}+P^-_1-Q$
we mean that $\tilde{\mathcal{D}}$ is an effective divisor
equivalent to 
$\mathcal{D}+P^-_1-Q.$

\noindent{\bf Proof.} Let us recall that after the Laplace transformation of
the first type the new $\psi$-function is 
$$
\tilde{\psi}_n=(1+v_nT)\psi_n=\psi_n+v_n\psi_{n+1}.
$$
But $\tilde{\psi}$ does not satisfy the normalization condition
$\psi_0(0)=1.$ Dividing by $\tilde{\psi}_0(0)$ we obtain
$$
\hat{\psi}_n(y)=\frac{\psi_n(y)+v_n(y)\psi_{n+1}(y)}{1+v_0(0)\psi_1(0)}.
$$
All terms in this formula can be expressed in terms of theta-functions
since $v_n=\frac{d_n}{b_n}.$ 
It follows from the consideration of poles and zeroes
that up to multiplication by a
constant $1+v_0(0)\psi_1(0)$ is equal to
$$
\exp(\int_{P_0}^P\!\!\!\Omega_{QP^-_1})%
\frac{\Theta(A(P)+A(Q)-A(P^-_1)-A(\mathcal{D})-\mathcal{K})}%
{\Theta(A(P)-A(\mathcal{D})-\mathcal{K})}.
$$
In the same way up to multiplication by a
constant $\psi_n(y)+v_n(y)\psi_{n+1}(y)$ 
is equal to
$$
\exp(\int_{P_0}^P\!\!\!\!\!y\Omega+\sumprime\limits_{i=1}^n%
\Omega_{P^+_iP^-_i}+\Omega_{QP^-_{n+1}})\times
$$
$$
\times\frac{\Theta(A(P)+yU+\sumprime\limits_{i=1}^n U_{P^+_iP^-_i}+%
A(Q)-A(P^-_{n+1})-A(\mathcal{D})-\mathcal{K})}%
{\Theta(A(P)-A(\mathcal{D})-\mathcal{K})}.
$$
We obtain that
$$
\hat{\psi}_n(y)=\hat{r}_n(y)%
\exp(\int_{P_0}^P\!\!\!\!\!y\Omega+%
\sumprime\limits_{i=1}^n\Omega_{P^+_iP^-_i}+%
\Omega_{QP^-_{n+1}}-\Omega_{QP^-_1})\times
$$
$$
\textstyle
\times\frac{\Theta(A(P)+yU+\sumprime\limits_{i=1}^nU_{P^+_iP^-_i}+%
A(Q)-A(P^-_{n+1})-A(Q)+A(P^-_1)-A(\mathcal{D})-\mathcal{K})}%
{\Theta(A(P)+A(Q)-A(P^-_1)-A(\mathcal{D})-\mathcal{K})},
$$
where  $\hat{r}_n(y)$ are constants.

Since
$
\sumprime\limits_{i=1}^n\Omega_{P^+_iP^-_i}+%
\Omega_{QP^-_{n+1}}-\Omega_{QP^-_1}=%
\sumprime\limits_{i=1}^n\Omega_{P^+_iP^-_{i+1}},
$
we obtain
$$
\hat{\psi}_n(y)=\hat{r}_n(y)%
\exp(\int_{P_0}^P\!\!\!\!\!y\Omega+%
\sumprime\limits_{i=1}^n\Omega_{P^+_iP^-_{i+1}})\times
$$
$$
\times\frac{\Theta(A(P)+yU+\sumprime\limits_{i=1}^nU_{P^+_iP^-_{i+1}}%
-A(\tilde{\mathcal{D}})-\mathcal{K})}%
{\Theta(A(P)-A(\tilde{\mathcal{D}})-\mathcal{K})},
$$
where $\tilde{\mathcal{D}}=\mathcal{D}+P^-_1-Q.$ This implies the
statement of this Theorem for the Laplace transformation of the first
type.

The formula for the Laplace transformations of the second type
can be easily obtained from the formula for the Laplace 
transformation of the first
type since they are inverse to each other. $\Box$

This Theorem makes it possible 
to easily construct chains of
Laplace transformations and, 
therefore, solutions of the equations~(\ref{eqw}) and~(\ref{2D1})
in terms of theta-functions.

\subsection{Spectral properties of the Laplace 
transformations of algebro-geometric two-dimensional 
discrete hyperbolic Schr\"odinger operators}

As was already explained in Section~\ref{discretelaplace},
the group generated by the Laplace transformations
has three generators. Let us now describe how these generators 
act on the spectral data.
\begin{Theorem}\label{laplsp}
The following identities hold:
\begin{align*}
\Lambda^{++}_{12}(P_i^-)=&P_{i+1}^-,&
\Lambda^{++}_{12}(P_i^+)=&P_{i}^+,&
\Lambda^{++}_{12}(Q_i^-)=&P_{i-1}^-,&
\Lambda^{++}_{12}(P_i^+)=&P_{i}^+,\\
S_1(Q_i^-)=&Q_{i-1}^-,& S_1(Q_i^+)=&Q_{i-1}^+,& S_1(P_i^-)=&P_{i}^-,&
S_1(P_i^-)=&P_{i}^+,\\
S_2(Q_i^-)=&Q_{i}^-,& S_2(Q_i^+)=&Q_{i}^+,& S_2(P_i^-)=&P_{i-1}^-,&
S_1(P_i^-)=&P_{i-1}^+,
\end{align*}
\begin{align*}
\Lambda^{++}_{12}(\mathcal D)&=\mathcal D+P_0^--Q^-_{-1},&
S_1(\mathcal D)&=\mathcal D+Q^+_{-1}-Q^-_{-1},
\end{align*}
\begin{align*}
S_2(\mathcal
D)&=\mathcal D+P^+_{-1}-P^-_{-1}.
\end{align*}
\end{Theorem}

\noindent{\bf Proof} is analogous to the Proof 
of Theorem~\ref{laplacesp}.
$\Box$

As it was mentioned in the Introduction,
cyclic chains of Laplace transformations were studied
in the paper~\cite{NV2} (we should also mention the paper~\cite{VS}
where cyclic chains of Darboux transformations were studied).

In our case we define a cyclic chain of Laplace transformations
as a chain such that $(\Lambda^{++}_{12})^\alpha(L)=S_1^\beta S_2^\gamma(L).$

Let us say that an operator $L$ is an integrable
operator if $P^\pm_i=P^\pm,$ $Q^\pm_i=Q^\pm.$
This condition means that the dynamics of the iterated
Laplace transformation is linearizable on the Jacobian of 
the spectral curve. Indeed,
it follows from the Theorem~\ref{laplsp} that in this case
the points $P^\pm,$ $Q^\pm$ are invariant and the divisor
$\mathcal{D}$ is shifted by a {\em fixed} vector $P^--Q^-.$

The condition of integrability is absolutely explicit in terms
of the coefficients of a periodic
operator $L.$ Indeed, it follows from the explicit formulas for the 
points $P^\pm_n$ and $Q^\pm_n$ that an operator $L$ is
integrable if and only if
$\frac{A_{.,j}}{B_{.,j}},$ $\frac{C_{.,j}}{D_{.,j}},$
$\frac{A_{i,.}}{C_{i,.}}$ and
$\frac{B_{i,.}}{D_{i,.}}$
are constants.

We have the following Theorem similar to
the one from  \cite{NV2}.
\begin{Theorem}
If $(\Lambda^{++}_{12})^\alpha(L)= S_1^\beta S_2^\gamma(L)$  and
$(\beta,\delta)=1$, $(\gamma,\epsilon)=1$,
$(\alpha+\gamma,\epsilon)=1$, $(\alpha-\beta,\delta)=1$, then
operator $L$ is an integrable operator.
\end{Theorem}

\noindent{\bf Proof} can be done by direct calculation. $\Box$

As in the semi-discrete case, we can construct solutions
of the completely discretized 2D Toda lattice~(\ref{completelydiscr}).
In this case we can do it for arbitrary generic periodic
initial data since any periodic operator is algebro-geometric.
This gives us the following Lemma.

\begin{Lemma}  The family of solutions
of the completely discretized 2D Toda 
lattice~(\ref{completelydiscr}) with a generic $\mathfrak{T}$-periodic
initial data $w_{n,m}^{(0)}$  can be written explicitly in terms of 
$\theta$-functions of the corresponding spectral curve.
This family is parameterized by a set of arbitrary $\mathfrak{T}$-periodic 
constants $H^{(0)}_{n,m}.$
\end{Lemma}

\section*{Acknowledgments}

One of the authors (A.~P.) is grateful to the Centre de Recherches
Math\'ematiques, Universit\'e de Montr\'eal, for its hospitality.
The authors thank Pavel Winternitz for useful discussions.

\end{document}